\documentclass[twocolumn]{aastex701}
\usepackage{amsmath}
\usepackage{xcolor}

\newcommand{\angstrom}{\textup{\AA}}

\begin{document}

\title{PAH Emission Spectra and Band Ratios for Arbitrary Radiation Fields\\ with the Single Photon Approximation}

\author[orcid=0000-0001-6325-9317]{Helena M. Richie}
\affiliation{Physics and Astronomy Department, University of Pittsburgh, 3941 O’Hara St, Pittsburgh, PA 15260, USA}
\email[show]{helenarichie@pitt.edu}  

\author[orcid=0000-0001-7449-4638]{Brandon S. Hensley} 
\affiliation{Jet Propulsion Laboratory, California Institute of Technology, 4800 Oak Grove Drive, Pasadena, CA 91109, USA}
\email{bhensley@jpl.nasa.gov}

\begin{abstract}

We present a new method for generating emission spectra from polycyclic aromatic hydrocarbons (PAHs) in arbitrary radiation fields. We utilize the single-photon limit for PAH heating and emission to treat individual photon absorptions as independent events. This allows the construction of a set of single-photon emission ``basis spectra'' that can be scaled to produce an output emission spectrum given any input heating spectrum. We find that this method produces agreement with PAH emission spectra computed accounting for multi-photon effects to within $\simeq10\%$ in the $3$--$20~{\rm \mu m}$ wavelength range for radiation fields with intensity $U<100$. We use this framework to explore the dependence of PAH band ratios on the radiation field spectrum across grain sizes, finding in particular a strong dependence of the 3.3 to 11.2\,$\mu$m band ratio on radiation field hardness. A Python-based tool and a set of basis spectra that can be used to generate these emission spectra are made publicly available.

\end{abstract}


\section{Introduction}

Polycyclic aromatic hydrocarbons (PAHs) are the carriers of prominent mid-infrared emission features frequently observed in galaxies, caused by the absorption of energy from starlight photons and subsequent vibrational emission between $3.3$--$17~{\rm \mu m}$ \citep{Leger1984, Allamandola1985, Tielens2008, Li2020}. The relative strengths of these emission features depend on the properties of the underlying PAH population, such as the size distribution and ionization function \citep[e.g.,][]{Maragkoudakis:2020, Draine2021}. 
In turn, these properties depend on the physical conditions of the interstellar medium (ISM) in which the PAHs reside, making PAH emission spectra a promising diagnostic of galaxy properties such as star formation rate \citep{Shipley:2016, Lai:2020, McKinney:2025}, hardness of the interstellar radiation field \citep{Draine2021, Rigopoulou:2021, Baron2025}, and metallicity \citep{Aniano:2020, Whitcomb:2024, Whitcomb:2025}.

The emission spectrum of a PAH depends upon its thermal history: the more time the grain spends at high temperatures, the more it radiates at short relative to long wavelengths. Indeed, observations have shown that the relative strengths of PAH emission features depend on the spectrum of the illuminating radiation field (\citealt{Donnelly2024, Baron2024, Baron2025}), as expected since UV photons can heat PAHs to higher temperatures than optical ones. However, this creates a degeneracy in observed band ratios between intrinsic PAH properties and the radiation field heating them. Disentangling this degeneracy requires quantification of how PAH emission spectra respond to changes in the interstellar radiation field. 

Stochastic heating of PAHs is a Markov process in which PAHs heat up from absorption of photons and cool down via vibrational emission. Some theoretical studies have modeled this process with a transition matrix between a discretized set of energy states (equivalently temperature states) from which the probability of a given PAH having a temperature $T$ in a given radiation field can be determined \citep{Guhathakurta:1989, Draine2001}. This method accounts for multi-photon processes in which a PAH absorbs another photon before cooling fully to its ground state. However, even optimized calculations can be slow owing to the quadratic scaling with the number of energy states. The importance of high-temperature states even with very small probability of occupation introduces numerical challenges as well. Other approaches based on Monte Carlo simulations \citep{Draine:1985} and iterative techniques \citep{Desert:1986} have even poorer scaling with the number of energy levels \citep[see discussion in][]{Guhathakurta:1989} and so are not a viable alternative without further algorithmic development.

The difficulty of computing PAH emission spectra has limited investigations into how changes in the spectrum of the interstellar radiation field alter the PAH spectrum. In their study of the effect of radiation field hardness on PAH band ratios, \citet{Rigopoulou:2021} employed monochromatic heating spectra without accounting for multi-photon effects. Recently, \citet{Draine2021} provided an extensive library of PAH emission spectra for a variety of radiation fields computed using the PAH model and stochastic heating framework of \citet{Draine2001}. While this has elucidated many of the effects of the radiation field spectrum on the PAH emission spectrum, ultimately the \citet{Draine2021} analysis and data products are restricted to a fixed set of radiation fields without a straightforward way to generalize to new ones. It is therefore the goal of this work to enable efficient computation of PAH emission spectra for arbitrary radiation fields.

Over a large range of strengths of the \citet{Mathis1983} interstellar radiation field, the $\sim$3--20\,$\mu$m PAH emission spectrum is nearly invariant in shape \citep{Draine2007}. This suggests that the emission is dominated by grains that cool completely before having a chance to absorb another photon. In this ``single photon'' limit, an emission spectrum resulting from an arbitrary radiation field can be computed if the emission spectrum resulting from the absorption of a photon of any given energy is known. We demonstrate that this ``single photon approximation'' yields accurate emission spectra compared to multi-photon calculations over a wide range of radiation field intensities and spectra while having negligible computational cost. This approach permits investigating PAH band ratios as a function of photon energy directly, limning the range of possible band ratios accessible to a specific PAH model. Software implementing the single photon approximation, as well as the accompanying ``basis spectra'' (i.e. spectra corresponding to individual photon absorptions) as a function of absorbed photon energy, are made publicly available.

This paper is organized as follows: in Section~\ref{sec:single-photon-approx}, we present the single photon approximation and its application to computing emission spectra for arbitrary heating spectra; in Section~\ref{sec:pah_model}, we detail the model of interstellar PAHs employed in this work; in Section~\ref{sec:validation}, we validate the single photon approximation against multi-photon calculations and quantify its regime of validity; in Section~\ref{sec:band_ratios}, we investigate PAH band ratios as a function of absorbed photon energy; in Section~\ref{sec:discussion}, we discuss these results in the context of current observations of PAH emission; in Section~\ref{sec:code}, we briefly describe the software and data products made available; and in Section~\ref{sec:conclusions}, we summarize our principal conclusions.

\section{The Single Photon Approximation} \label{sec:single-photon-approx}

The primary source of PAH heating is stochastic absorption of starlight photons, predominantly in the UV \citep{Leger1984, Allamandola1985}. 
The energy absorbed from these photons briefly raises the internal temperature of grains and is eventually re-radiated primarily in the form of broad mid-infrared emission features. 
The rate of photon absorption is determined in part by the absorption cross-section, $C_\mathrm{abs}$, which depends on the dust grain size, shape, ionization state, and the photon wavelength. 
Larger grains have a larger geometric cross-section and generally absorb photons at a higher rate than smaller grains. 
In the large grain limit ($a\gtrsim50~$\AA, where $a$ is the effective radius), photon absorptions occur frequently compared to the grain cooling time, and each photon imparts an energy small enough compared to the total internal energy of the grain, such that the temperatures of large grains can be approximated as steady-state.
In the diffuse ISM, the steady-state temperature of large grains tends to be $\sim20~\textrm{K}$.

\begin{figure*}
\begin{centering}
\includegraphics[width=\textwidth]{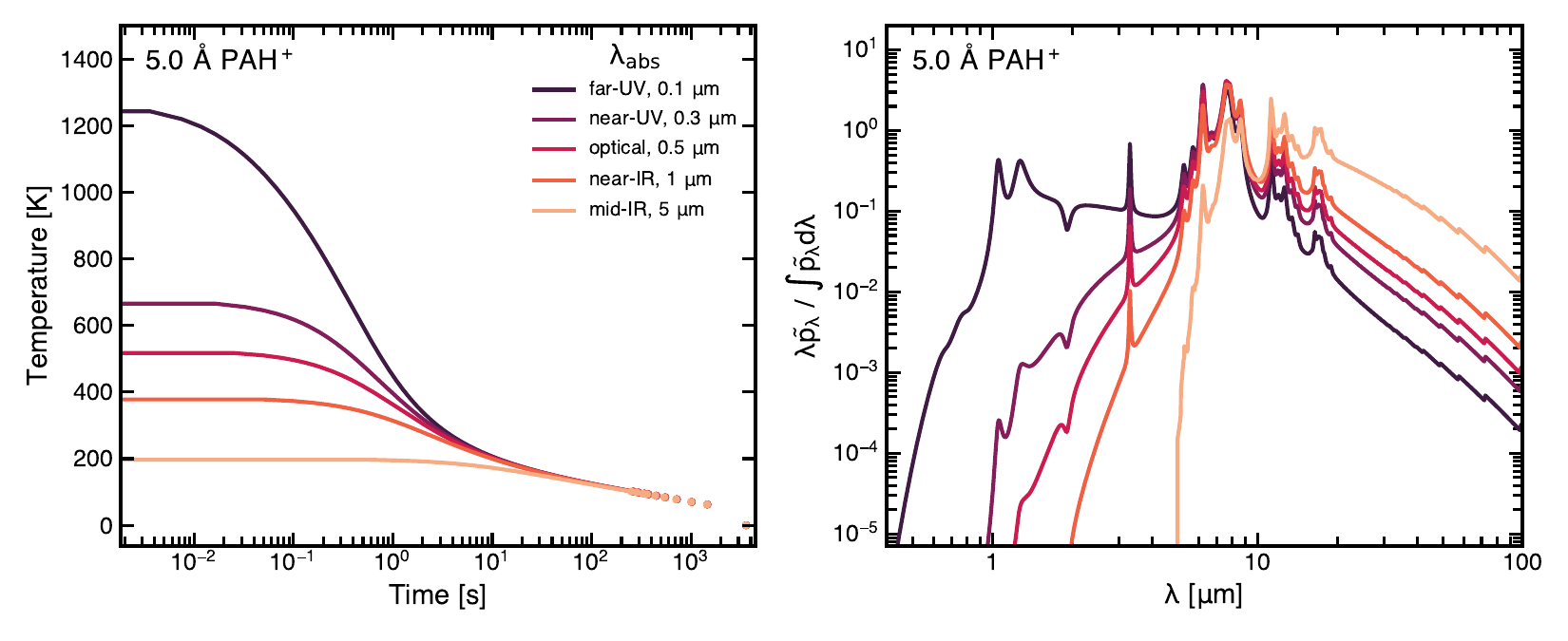}
\caption{(Left) temperature evolution of a $5~\angstrom$ ionized PAH as it radiates the energy absorbed from individual photons ranging from far-UV to mid-infrared wavelengths. The dots indicate the discrete energy (equivalently, temperature) bins discussed in Section~\ref{subsec:energy_binning}. (Right) the resulting emission spectra, $\tilde{p}_{\lambda_{\rm em}}(\lambda_{\rm abs})$, which we adopt as our basis spectra.}\label{fig:basisvectors}
\end{centering}
\end{figure*}

Small grains have a smaller $C_\mathrm{abs}$ and a lower heat capacity than large grains, and so undergo a large temperature change when they absorb photons. 
Indeed, a single photon can briefly raise a small grain's temperature to hundreds of Kelvin or more. 
Across a wide range of radiation fields, the time between photon absorptions for small grains is typically longer than the time it takes for a grain to cool back down to its ground state \citep{Draine2001}.
When this condition holds, a grain is in the ``single-photon limit," where individual photon absorptions occur independently of one another.
Because of this effect, the resulting infrared emission features from these grains, particularly in the $\sim$1--$20~\mu$m range, can be written as a linear combination of the emission spectra from individual photon absorption events.

For a dust grain of a given size and composition, the power radiated per unit wavelength $\tilde{p}_{\lambda_{em}}({\lambda_{\rm abs})}$ following the absorption of a single photon of wavelength $\lambda_{\rm abs}$ can be determined by computing the time evolution of the grain temperature. 
The energy absorbed by the grain is

\begin{equation}
    E_\mathrm{abs}=\frac{h c}{\lambda_\mathrm{abs}},
    \label{eq:energy_init}
\end{equation}
such that the internal energy $E$ of the grain at time $t = 0$ is $E_{\rm abs}$.
In the continuous cooling limit (see \citealt{Draine2001}), the grain cools radiatively according to 

\begin{equation}
    \frac{dE}{dt}=-\int_{\lambda_{\rm abs}}^\infty 4\pi \, B_\lambda(T(t))\,C_\mathrm{abs}(\lambda)\,d\lambda,
    \label{eq:cooling_rate}
\end{equation}

\noindent where $B_\lambda$ is the Planck function.
Following \cite{Draine2001}, we restrict the wavelengths of emitted photons $\lambda_{\rm em}$ to $\lambda_{\rm em}\geq\lambda_{\rm abs}$ so the PAH cannot emit at higher energies than the absorbed photon energy.
Given a model for the internal energy of a grain as a function of grain temperature $T$, i.e., $E(T)$, one can solve Equation~\ref{eq:cooling_rate} with the boundary condition defined by Equation~\ref{eq:energy_init} to obtain the grain temperature evolution $T(t)$ as it cools to $T\rightarrow0$.
Example $T(t)$ are shown in the left column of Figure~\ref{fig:basisvectors} for a $5~\angstrom$ ionized PAH over a range of $\lambda_{\rm abs}$.

The temperature evolution $T(t)$ determines the relative power $\tilde{p}$ emitted at each wavelength $\lambda_{\rm em}$ during the cooling process:

\begin{align}
\begin{aligned}
&\tilde{p}_{\lambda_\mathrm{em}}(\lambda_{\rm abs}) = \\
&\begin{cases}
    0 & \text{for } \lambda_{\rm em} < \lambda_{\rm abs} \\
    \frac{4\pi}{t_{\rm max}}\int_0^{t_{\rm max}} B_{\lambda_{\rm em}}(T(t))\, C_\mathrm{abs}(\lambda_{\rm em})\,dt  & \text{for } \lambda_{\rm em} \geq\lambda_{\rm abs}.
\end{cases}
\label{eq:basisvector}
\end{aligned}
\end{align}

\noindent As in Equation~\ref{eq:cooling_rate}, we restrict the emission wavelengths so that PAH emission energies do not exceed $E_{\rm abs}$.\footnote{This results in the presence of prominent ``sawtooth" spectral features in the far-infrared of $\tilde{p}_{\lambda_{\rm em}}(\lambda_{\rm abs})$ (visible in Figure~\ref{fig:basisvectors}), similar to those discussed in \citet{Draine2001}.} This $\tilde{p}_{\lambda_\mathrm{em}}$ can be thought of as a basis spectrum of the emission spectrum $p_{\lambda_\mathrm{em}}$, where the basis spectrum is the corresponding emission from a given grain after absorbing a photon of energy $E_\mathrm{abs}$.
Since this methodology requires only that $\tilde{p}_{\lambda_{\rm em}}$ is defined such that the power emitted at the various $\lambda_{\rm em}$ be normalized correctly relative to each other, the prefactor in Equation~\ref{eq:basisvector} is arbitrary.
Examples of these basis spectra are shown in the right column of Figure~\ref{fig:basisvectors} for a range of absorbed photon wavelengths. For alternate derivations of the single-photon emission spectrum, see \citet{Leger:1989} and \citet{Rigopoulou:2021}.


Consider a grain heated by a radiation field $u_\lambda$, where $u_\lambda d\lambda$ is the energy density of photons having wavelength between $\lambda$ and $\lambda+d\lambda$. The amount of energy the grain absorbs from photons of energy $\lambda_{\rm abs}$ must be balanced by the energy radiated with spectrum described by $\tilde{p}_{\rm \lambda_{em}}(\lambda_{\rm abs})$. This energy balance sets the normalization required to compute the power emitted at each wavelength $\lambda_{\rm em}$, $p_{\lambda_{\rm em}}\left(\lambda_{\rm abs}\right)$.

A complication is that $\tilde{p}_{\rm \lambda_{em}}(\lambda_{\rm abs})$ can be computed only at a finite set of $\lambda_{\rm abs}$. Therefore, we assume that $\tilde{p}_{\rm \lambda_{em}}(\lambda_{\rm abs})$ is constant over a range of wavelengths between $\lambda_{\rm abs}$ and $\lambda_{\rm abs}\left(1+\Delta\right)$, where $\Delta$ is the dimensionless fractional width in wavelength range. We compute $\tilde{p}_{\rm \lambda_{em}}(\lambda_{\rm abs})$ in 474 wavelength bins between 912\,\AA\ and 10.1\,$\mu$m having $\Delta = 0.01$.

With this formulation,

\begin{equation}
    p_{{\lambda_\mathrm{em}}}(\lambda_\mathrm{abs})=\frac{\int_{\lambda_\mathrm{abs}}^{\lambda_\mathrm{abs}(1+\Delta)}cu_\lambda(\lambda) C_\mathrm{abs}(\lambda)d\lambda}{\int_{\rm \lambda_{em}} \tilde{p}_{\lambda_\mathrm{em}}(\lambda_\mathrm{abs})d\lambda}\tilde{p}_{{\lambda_\mathrm{em}}}(\lambda_\mathrm{abs}).
\label{eq:scaled_basis_vector}
\end{equation}

\noindent Given a set of basis spectra that spans the entire wavelength range over which photon absorption occurs, the sum of these $p_{\lambda_\mathrm{em}}(\lambda_\mathrm{abs})$ over all $\lambda_{\rm abs}$, i.e.,

\begin{equation}
    p_{\rm \lambda_{em}}=\sum_{\lambda_{\rm abs}}p_{{\lambda_\mathrm{em}}}(\lambda_\mathrm{abs})
\label{eq:integrated_spectrum}
\end{equation}

\noindent yields the full emission spectrum for that grain in the specified radiation field in the single-photon limit.

\section{PAH Model} \label{sec:pah_model}
The method described in Section~\ref{sec:single-photon-approx} is independent of the grain type, and could in principle be applied for an arbitrary dust model. In this Section, we describe the PAH model used in this work.

\subsection{Material Properties}
We largely follow the PAH model described in \citet{Draine2001}, with updates following subsequent publications (\citealt{Draine2007}; \citealt{Draine:2016}; \citealt{Draine2021}). For completeness, we provide a self-contained description of the model here.

Let $N_{\rm C}$ and $N_{\rm H}$ denote the number of carbon and hydrogen atoms in a PAH molecule, respectively.
All PAHs in the model are assumed to have a mass density of $2.0~{\rm g\, cm^{-3}}$ \citep{Draine2021}. 
Therefore, a PAH molecule with an effective radius\footnote{Defined as the radius of a sphere having equal volume} of $a$ contains

\begin{equation}
    N_{\rm C}=418\left( \frac{a}{10~\angstrom} \right) ^3
\end{equation}

\noindent carbon atoms with $N_{\rm C}$ rounded to the nearest integer. The number of hydrogen atoms is approximated as

\begin{equation}
    N_\mathrm{H} = \begin{cases}
    0.5N_\mathrm{C}+0.5 & \text{for } N_\mathrm{C}\leq25, \\
    2.5\sqrt{N_\mathrm{C}} + 0.5 & \text{for } 25<N_\mathrm{C}\leq100, \\
    0.25N_\mathrm{C}+0.5 & \text{for } N_\mathrm{C}>100,
    \end{cases}
\end{equation}

\noindent also rounded to the nearest integer.

The internal energy $E$ of a PAH molecule is approximated as the sum of the energy contributions from its individual vibrational degrees of freedom. 
A PAH molecule with $N_\mathrm{C}$ carbon atoms and $N_\mathrm{H}$ hydrogen atoms has $N^m_\mathrm{tot}=3(N_\mathrm{H}+N_\mathrm{C}-2)$ vibrational modes from C--C and C--H in-plane (``ip'') and out-of-plane (``op'') bending and C--H stretching (``str''). 
The modes are approximated as harmonic oscillators with fundamental frequency $\omega_j$, for which the expectation value for the energy as a function of temperature $T$ is

\begin{equation}
    E(T)=\sum_{j=1}^{N^m_\mathrm{tot}}\frac{\hbar\omega_j}{\exp(\hbar \omega_j/kT)-1},
    \label{eq:mode_energy}
\end{equation}

\noindent assuming thermal equilibrium. 

The mode frequencies for most PAH molecules are unknown, so they must be approximated. The C-C modes are modeled with a two-dimensional Debye spectrum with Debye temperatures of $\Theta_{\rm op,CC}=863~\textrm{K}$ and $\Theta_{\rm ip,CC}=2500~\textrm{K}$ \citep{Krumhansl1953}. The C-C mode frequencies are given by

\begin{equation}
    \hbar \omega_{j, {\rm CC}}=k\Theta_{\rm CC}~\bigg[ \frac{1-\beta}{N^m_{\rm CC}}(j-\delta_j)+\beta \bigg]^{1/2}.
    \label{eq:cc_mode_frequency}
\end{equation}

\noindent Here, $\delta_j$ adjusts the second and third mode frequencies to align with the observed frequencies of coronene (see \citealt{Draine2001} for details),

\begin{equation}
    \delta_j=\begin{cases}
    \frac{1}{2} & \text{for }j \ne 2 \text{ or }3, \\
    1  & \text{for } j=2 \text{ or } j=3
\end{cases}
\end{equation}

\noindent and $\beta$ depends on the shape of the molecule,

\begin{equation}
    \beta=\begin{cases}
        0 & \text{for } N_\mathrm{C}\leq 54, \\
        \frac{1}{2N^m_{\rm CC}-1}\Big(\frac{N_\mathrm{C}-54}{52}\Big) & \text{for } 54 < N_\mathrm{C} \leq 102, \\
        \frac{1}{2N^m_{\rm CC}-1}\Big[ \frac{N_\mathrm{C}-2}{52}\Big(\frac{102}{N_\mathrm{C}}\Big)^{2/3}-1 \Big] & \text{for } N_\mathrm{C} > 102
    \end{cases}
\end{equation}

\noindent where $N_\mathrm{C}\leq54$ is the planar case and $N_\mathrm{C}\geq 102$ is the spherical case. 
The number of modes for C--C in-plane bending is $N^m_{\rm ip,CC}=N_\mathrm{C}-2$, and $N^m_{\rm op, CC}=2(N_\mathrm{C}-2)$ for out-of-plane bending.

The adopted C--H vibrational frequencies are

\begin{equation}
    \hbar\omega_{j, {\rm CH}}=k\Theta_{\rm CH},
\end{equation}

\noindent where $\Theta_\mathrm{op,CH}=1257~\textrm{K}$, $\Theta_\mathrm{ip,CH}=1670~\textrm{K}$, and $\Theta_\mathrm{str,CH}=4360~\textrm{K}$, corresponding to frequencies of $11~{\mu\textrm{m}^{-1}}$, $8.6~{\mu\textrm{m}^{-1}}$, and $3.3~{\mu\textrm{m}}^{-1}$, respectively. 
For each of these three C--H modes, there are $N^m_{\rm CH}=N_\mathrm{H}$ contributions to Equation~\ref{eq:mode_energy}.

For large PAHs, the number of modes can become intractable. 
Therefore, the method described above to calculate $E(T)$ is used only for grains having $N_\mathrm{C}\leq7360$. 
For larger grains, the contributions from the C--C modes in Equation~\ref{eq:mode_energy} are replaced with the continuum Debye model:

\begin{align}
    E_{\rm PAH}^{\rm CC} &= 2\left(N_{\rm C}-2\right)\left[k\Theta_{\rm op,CC}\left(\frac{T}{\Theta_{\rm op,CC}}\right)^3 \int_0^{\Theta_{\rm op,CC}/T} \frac{u^2 du}{e^u-1} \right. \nonumber\\  
    & \left. + 2 k\Theta_{\rm ip,CC}\left(\frac{T}{\Theta_{\rm ip,CC}}\right)^3 \int_0^{\Theta_{\rm ip,CC}/T} \frac{u^2 du}{e^u-1}\right]\,, 
\label{eq:debye_energy}\end{align}

\noindent which is equivalent to Equation~33 of \citet{Draine2001}\footnote{Equation~10 of \citet{Draine2001}, upon which their Equation~33 depends, has a typographical error: the prefactor in front of the integral should be $n$, not $1/n$. None of their calculations were affected}. 

We employ the PAH absorption cross-sections $C_{\rm abs}$ described in \citet{Draine2021}\footnote{Note erratum \citep{Draine2025}}. The cross sections depend on size, wavelength, and ionization state, where the latter is implemented as a binary parameter in which each PAH is either ``neutral'' (denoted ${\rm PAH^0}$) or ``ionized'' (denoted ${\rm PAH^+}$). 

\subsection{Energy Binning} \label{subsec:energy_binning}
To solve for the PAH cooling function $T(t)$, we employ the thermal continuous approximation (see \citealt{Draine2001} for details).
We iteratively decrease the PAH temperature as it loses energy according to the cooling rate (Equation~\ref{eq:cooling_rate}).
To integrate the cooling rate and evolve the PAH temperature, we divide the energy range $[0, E(\lambda_{\rm abs})]$ into discrete bins of width $dE$.
At high energies, the width of these bins is adaptively set to be a constant fraction of the PAH internal energy at the start of each timestep (we find convergence in the solution for $dE/E\leq0.03$).

At low energies, quantization of the vibrational energy levels cannot be ignored. Therefore, we follow the method described in Appendix~B of \citealt{Draine2001} to discretize the ten lowest energy bins.
Once the PAH cools to an energy of $\frac{1}{2}(\hbar\omega_{18}+\hbar\omega_{19})$, where $\omega_j=1,...,3(N_a-2)$ are the fundamental vibrational modes in order of increasing frequency, we define the bin edges as, 

\begin{equation}
    E_{1,\rm min}=\frac{3}2\hbar \omega_1-\frac{1}2\hbar\omega_2, \\
\end{equation}
\begin{equation}
    E_{j,\rm max}=E_{j+1,\rm min}=\frac{1}2(\hbar \omega_j+\hbar\omega_{j+1}) \text{ for } j=1,2,
\end{equation}
\begin{equation}
    E_{j,\rm max}=E_{j+1,\rm min}=\frac{1}2(\hbar\omega_{2j-2}+\hbar\omega_{2j-1}) \text{ for } 3\leq j\leq10.
\end{equation}

\noindent With these definitions, bins 1 and 2 contain the first two vibrational modes, and bins 3--10 each include two modes. The lowest energy bin is defined as $[0, E_{1,\rm min}]$.
An example of these bins for a $5\,\angstrom$ ionized PAH is shown in Figure~\ref{fig:basisvectors}.

\section{Validation} \label{sec:validation}

In this Section, we quantify the accuracy of the single photon approximation (hereafter SPA) by comparing spectra computed with our framework to the calculations of \citet{Draine2021} that account for the effects of multi-photon heating. 
We first consider the emission spectra of individual grains and then emission spectra integrated over a grain size distribution.
In all cases, we assess agreement as a function of both radiation field intensity and spectral shape.

\subsection{Multi-photon Effects on the PAH Emission Spectrum}
\label{subsec:multi-photon}

\begin{figure*}
\begin{centering}
\includegraphics[width=0.9\textwidth]{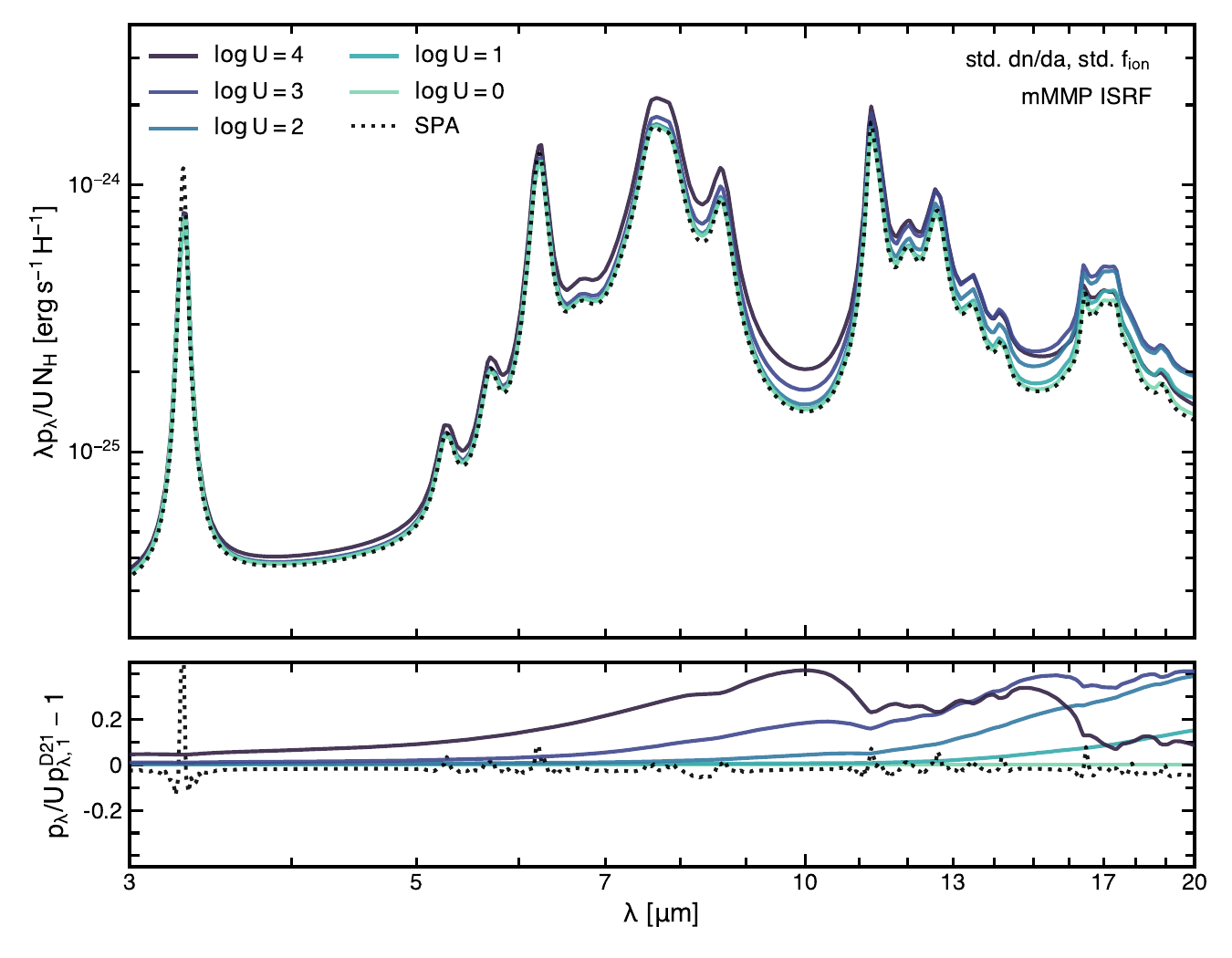}
\caption{The size- and ionization-integrated spectra for PAHs in the mMMP radiation field, normalized by $U$. We use the standard size distribution and ionization function defined in \citet{Draine2021}. The black dotted line shows the SPA spectrum, which has no $U$-dependence after normalization. The solid colored lines show models from \citet{Draine2021}, with line color indicating the value of $U$. The bottom panel shows the \citet{Draine2021} and SPA spectra for each $U$ normalized by $U p_{\lambda,1}$, where $p_{\lambda,1}$ is the $U=1$ spectrum.}\label{fig:size_integrated_spectrum}
\end{centering}
\end{figure*}

The fundamental assumption of the SPA is that upon absorbing a photon, the grain has time to cool completely before absorbing another. There are two primary consequences of multi-photon effects that are neglected by this approximation. First, absorption of additional photons when the grain has non-zero internal energy means that the grain spends more time at higher temperatures and less time at lower temperatures, shifting power in the emission spectrum from longer wavelengths toward shorter wavelengths. Second, it becomes possible for a grain to attain a higher internal energy than that of the highest energy photon absorbed. Even if rare, this effect can dramatically change the emission spectrum at short near-infrared wavelengths ($\lambda \lesssim 2\,\mu$m).

To quantify the magnitude of multi-photon effects, Figure~\ref{fig:size_integrated_spectrum} presents PAH emission spectra from \citet{Draine2021} (acquired from their publicly-available data repository, \citealt{DraineData2021}) for a range of radiation field intensities. The PAH model employed uses the standard grain size distribution and ionization function from that work. The radiation field is the modified \citet{Mathis1983} (mMMP) radiation field $u_{\rm \lambda,mMMP}$ (see \citealt{Draine2011} for details) scaled by a dimensionless factor quantifying the overall radiation field intensity, $U\equiv u/u_{\rm mMMP}$ (where $u$ is the integrated energy density of the radiation field), ranging from 1 to $10^4$. As evident from Equation~\ref{eq:scaled_basis_vector}, scaling the radiation field by a factor of $U$ under the SPA simply scales the emission spectrum by the same factor. Thus, differences in the spectral shapes of the emission spectra as a function of $U$ with multi-photon calculations illustrate the impact of multi-photon absorption on the emission spectrum.

The fractional differences presented in Figure~\ref{fig:size_integrated_spectrum} demonstrate that multi-photon effects make no more than a 10\% difference to the $\lambda \lesssim 10\,\mu$m emission spectrum for $U \lesssim 100$. For larger values of $U$ and longer wavelengths, the discrepancies become more severe. While the observed shift in power from long to short wavelengths as $U$ increases is consistent with expectations from multi-photon effects, it is possible that differences in energy binning as a function of $U$ in the \citet{Draine2001} framework also contribute. Thus, these differences are likely an upper bound on the importance of multi-photon effects. We treat them as the limiting accuracy of the SPA.

\subsection{Single Grain Spectra} \label{subsec:single-grain-spectra}

\begin{figure*}
\begin{centering}
\includegraphics[width=\textwidth]{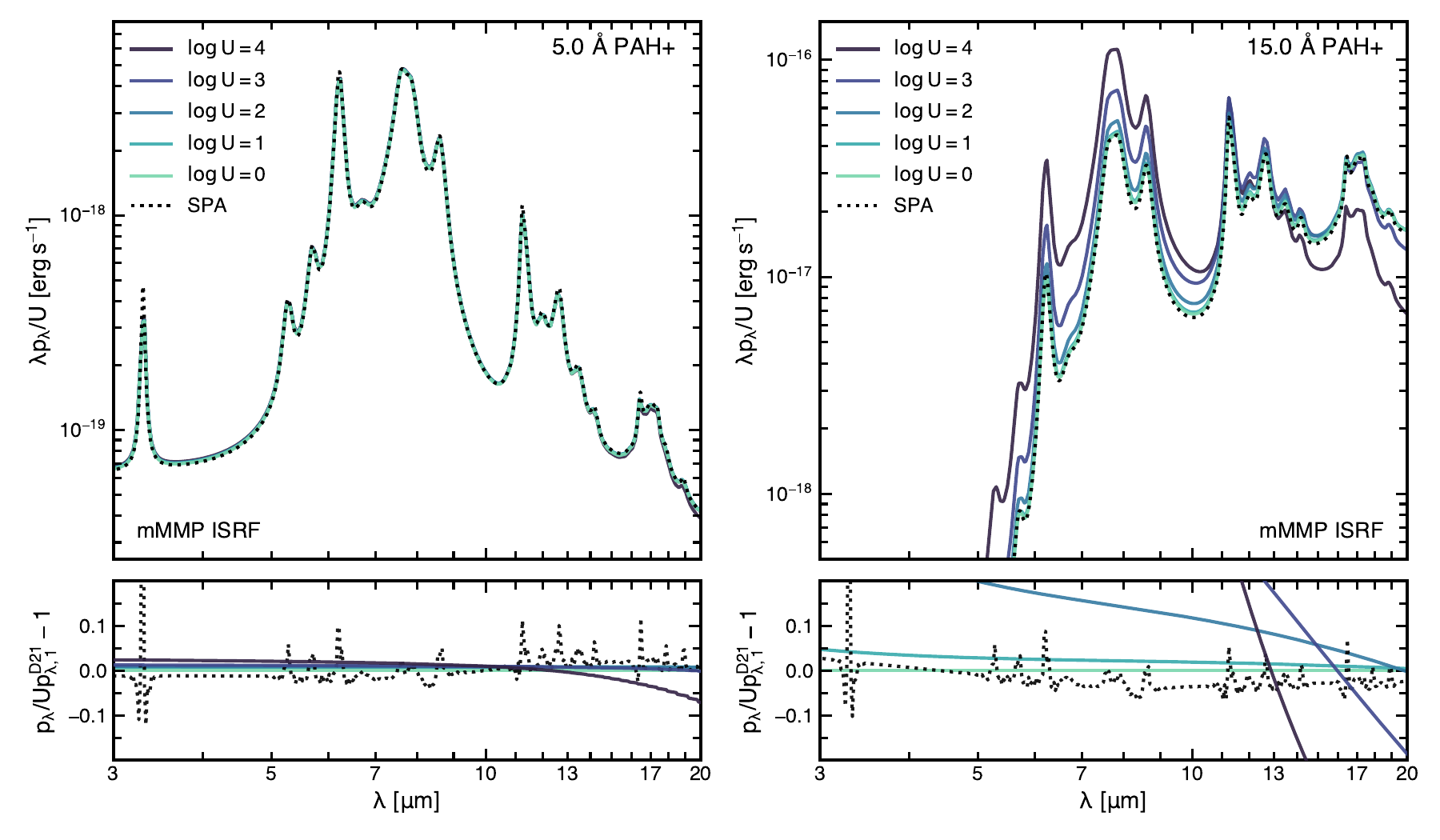}
\caption{Same as Figure~\ref{fig:size_integrated_spectrum}, but for individual grains. In the left and right panels we show results for $a=5.0~\angstrom$ and $15.0~\angstrom$ ionized PAHs, respectively.}\label{fig:single_grain_spectrum}
\end{centering}
\end{figure*}

We continue the comparison of the \citet{Draine2021} models as a function of $U$ in Figure~\ref{fig:single_grain_spectrum}, here for single $a=5$ and 15\,\AA\ ionized PAHs. As in Figure~\ref{fig:size_integrated_spectrum}, we normalize by $U$ to highlight the multi-photon effects.

The 5\,\AA\ PAH spectrum has little variation as $U$ is increased from 1 to $10^4$. Such a small grain absorbs photons infrequently, and so remains in the single photon limit even in intense radiation fields. The $U=10^4$ departs from the $U=1$ spectrum at only the 5\% level at 20\,$\mu$m, and much less at shorter wavelengths. Hence, we expect the SPA to be an excellent approximation for this grain.

In contrast, the 15\,\AA\ PAH spectra diverge sharply as $U$ is increased. While $U=1$ and $U=10$ agree to better than 5\% at $3 < \lambda/\mu{\rm m} < 20$, the $U=100$ spectrum differs by more than 20\% over most of that range. Clearly the SPA is a poor approximation for a 15\,\AA\ PAH once the radiation field much exceeds $U=10$. Given the fairly good agreement between the $U=1$ and $U=100$ spectra when integrated over a fiducial grain size distribution (Figure~\ref{fig:size_integrated_spectrum}), the 15\,\AA\ PAHs must contribute modestly to the total emission at $\lambda \lesssim 10\,\mu$m.

\begin{figure}
\begin{centering}
\includegraphics[width=0.48\textwidth]{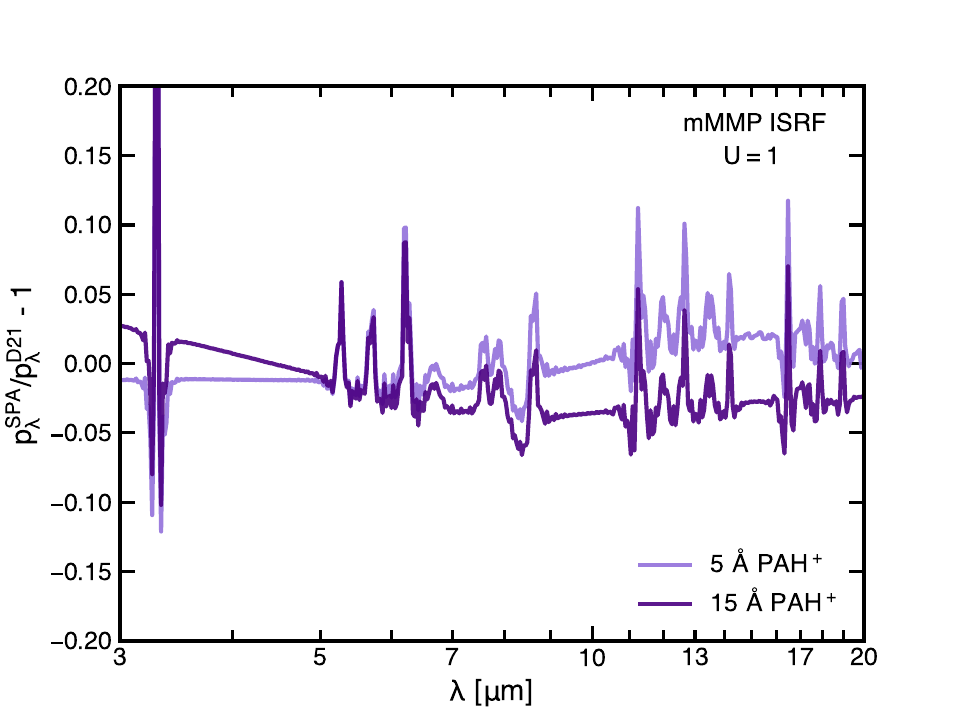}
\caption{Fractional residuals between single-grain spectra generated using the SPA ($p_{\lambda}^{\rm SPA}$) and from \citet{Draine2021} ($p_{\lambda}^{\rm D21}$) for grains heated by the mMMP radiation field with $U=1$. The sharp features are caused by interpolation effects near the peaks of the PAH emission features in the \citet{Draine2021} spectra, whereas overall offsets may arise from numerical differences between implementations.}\label{fig:residuals}
\end{centering}
\end{figure}

Figure~\ref{fig:residuals} presents a detailed comparison between the $U=1$ \citet{Draine2021} spectra of 5 and 15\,\AA\ PAHs and the corresponding spectra computed with our SPA framework (equivalent to the SPA residuals shown in Figure~\ref{fig:single_grain_spectrum}). Because multi-photon effects have been demonstrated to be modest at this low $U$ value, this figure principally highlights disagreements between models. We find that the \citet{Draine2021} spectra are reproduced by our SPA approach to within 3\% accuracy between 3 and 20\,$\mu$m.

Striking in Figure~\ref{fig:residuals} are the sharp features near the peaks of the PAH emission features. These are artifacts of the interpolation procedure used to compute the \citet{Draine2021} spectra: the wavelength sampling of $C_{\rm abs}$ is insufficient to resolve the features, and so large relative errors can be produced near wavelengths where the spectrum is changing rapidly. Because of the computational efficiency of the SPA framework, we employ $C_{\rm abs}$ at very high spectral resolution to fully resolve the PAH features and correct these errors.

Aside from these interpolation artifacts, some residual differences between models remain. Ultimately, the SPA results presented here and the \citet{Draine2021} spectra arise from different fundamental calculations: explicit calculation of a cooling curve in the former case, and solving a transition matrix between energy states in the latter. Small numerical discrepancies could arise purely from the differences in these approaches, such as from relatively coarse discretization of the energy levels in the \citet{Draine2001} method. The $\lesssim 3\%$ residuals originating from model differences are generally much smaller than the $\sim$10\% accuracy of the SPA itself vis-a-vis a full calculation with multi-photon effects (see Section~\ref{subsec:multi-photon}).

\subsection{Size-integrated Spectra} \label{subsec:size-integrated-spectrum}

\begin{figure*}
\begin{centering}
\includegraphics[width=\textwidth]{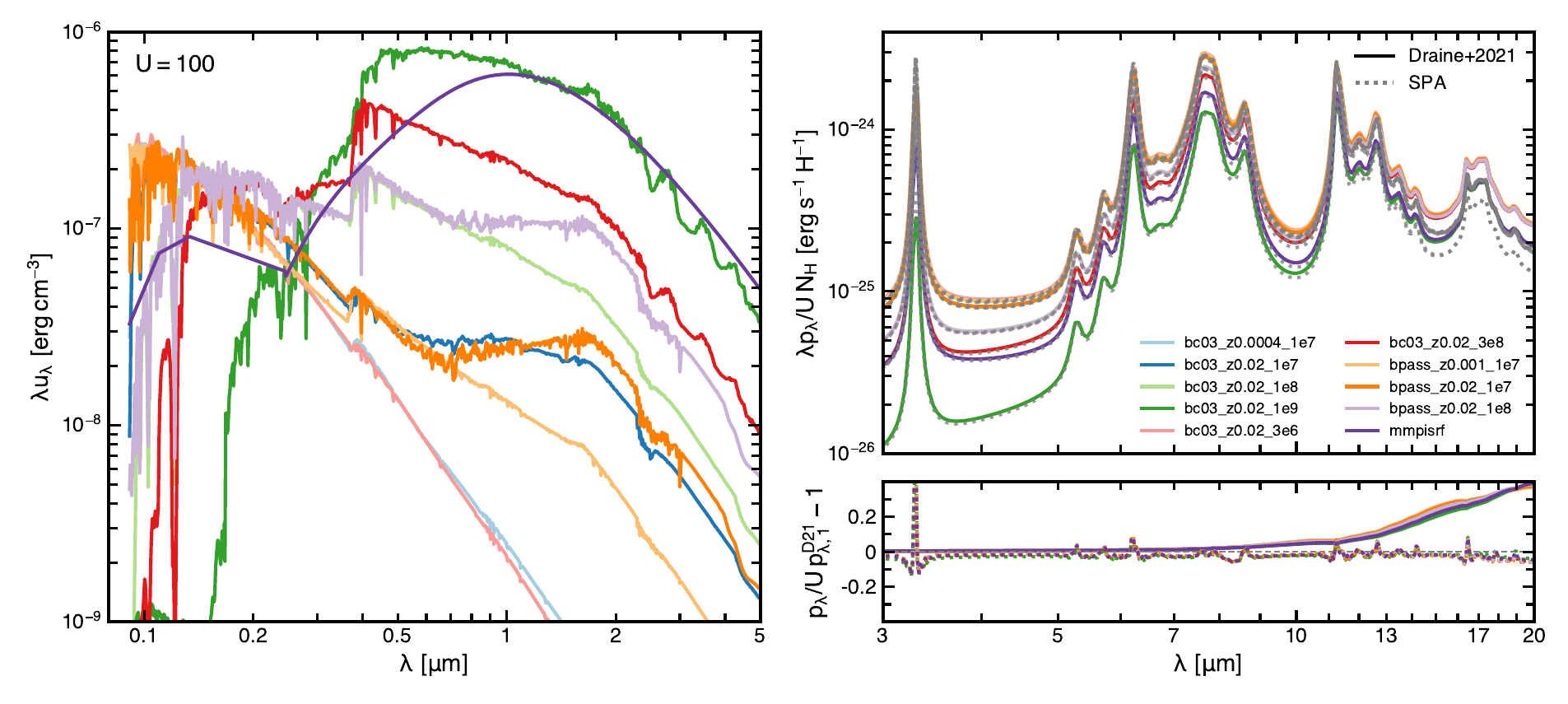}
\caption{(Left) ten radiation field models from \citet{Draine2021}, all with $U=100$. (Right) The resulting emission spectra, integrated over the standard size distribution and ionization function. Dotted lines show the SPA models and solid lines show the \citet{Draine2021} models. Note that some spectra are not readily distinguishable from each other on this scale. The bottom right panel shows the \citet{Draine2021} and SPA spectra normalized by $100\, p_{\lambda,1}$, where $p_{\lambda,1}$ is the $U=1$ spectrum.} \label{fig:rad_fields}
\end{centering}
\end{figure*}

Figure~\ref{fig:single_grain_spectrum} illustrates that the validity of the SPA is a strong function of grain size. We therefore turn our focus to size-integrated spectra to assess the accuracy of the SPA for standard PAH size distributions.

In addition to the suite of \citet{Draine2021} spectra, Figure~\ref{fig:size_integrated_spectrum} presents the SPA spectrum computed with our framework. 
The SPA spectrum agrees well with the \citet{Draine2021} spectra across the entire wavelength range of interest for $U=1-10$.
As discussed in Section~\ref{subsec:single-grain-spectra}, even 15\,\AA\ PAHs in the $U=1-10$ mMMP are primarily in the single-photon limit at these intensities, and introduce discrepancies to the spectra of $\lesssim5\%$ in the $3-20~{\rm \mu m}$ range. 
For $U\gtrsim100$, multi-photon effects can introduce $>10\%$ discrepancies at long wavelengths, which become increasingly dominated by large grains for which the SPA is poor.
On the other hand, the small grains that dominate the spectrum at short wavelengths remain in the single-photon limit across a wide range of $U$, so even the $U=10^4$ \citet{Draine2021} spectrum agrees with the SPA spectrum below $\sim$6$~{\rm \mu m}$.
As a result, the SPA model is consistently successful in reproducing the $3.3~{\rm \mu m}$ feature.

In Figure~\ref{fig:rad_fields}, we show size- and ionization-integrated SPA and \citet{Draine2021} spectra for a diverse set of radiation fields, all with $U=100$. 
Figure~\ref{fig:rad_fields} also compares the differences between $U=1$ and $U=100$ \citet{Draine2021} spectra to the differences between the $U=1$ and SPA spectra. 
For all of the radiation environments considered, the SPA model reproduces the $U=1$ \citet{Draine2021} spectra to excellent ($<5\%$) agreement across the entire $3-20~{\rm \mu m}$ range.
The consistency in shape of these residuals indicates that the small disagreement between our SPA calculations and the $U=1$ \citet{Draine2021} spectra is not a strong function of the shape of the radiation field.

The $U=100$ models systematically differ from the $U=1$ spectra due to multi-photon effects.  However, there is little difference among the residuals as a function of radiation spectrum.
This close agreement illustrates that the impact of multi-photon heating on the emission spectrum does not strongly depend on the shape of the heating spectrum.
Thus, the conclusion that the SPA provides an adequate description of the $3$--$20\,\mu$m PAH emission spectrum for $U \lesssim 100$ should hold across a large range of radiative environments.

\subsection{Validation Summary}

We have demonstrated that the validity of the SPA depends little on heating spectrum but strongly on the grain size distribution and the wavelength range modeled. We find that our SPA spectra agree with more extensive multi-photon calculations to within 5\% for $\lambda < 20\,\mu$m in $U=1$ radiation fields, regardless of spectral shape. For $U \leq 100$, they agree to within $\simeq$10\% for $\lambda < 10\,\mu$m, and to within 30\% for $\lambda < 20\,\mu$m with the fiducial PAH size distribution. 10\% accuracy is achieved at $\lambda < 6\,\mu$m for $U \leq 10^4$. Caution should be exercised when using SPA spectra for $U > 100$, $\lambda > 20\,\mu$m, or size distributions favoring large PAHs.

\section{PAH Band Ratios in the Single Photon Limit} \label{sec:band_ratios}

\begin{figure*}
\begin{centering}
\includegraphics[width=0.9\textwidth]{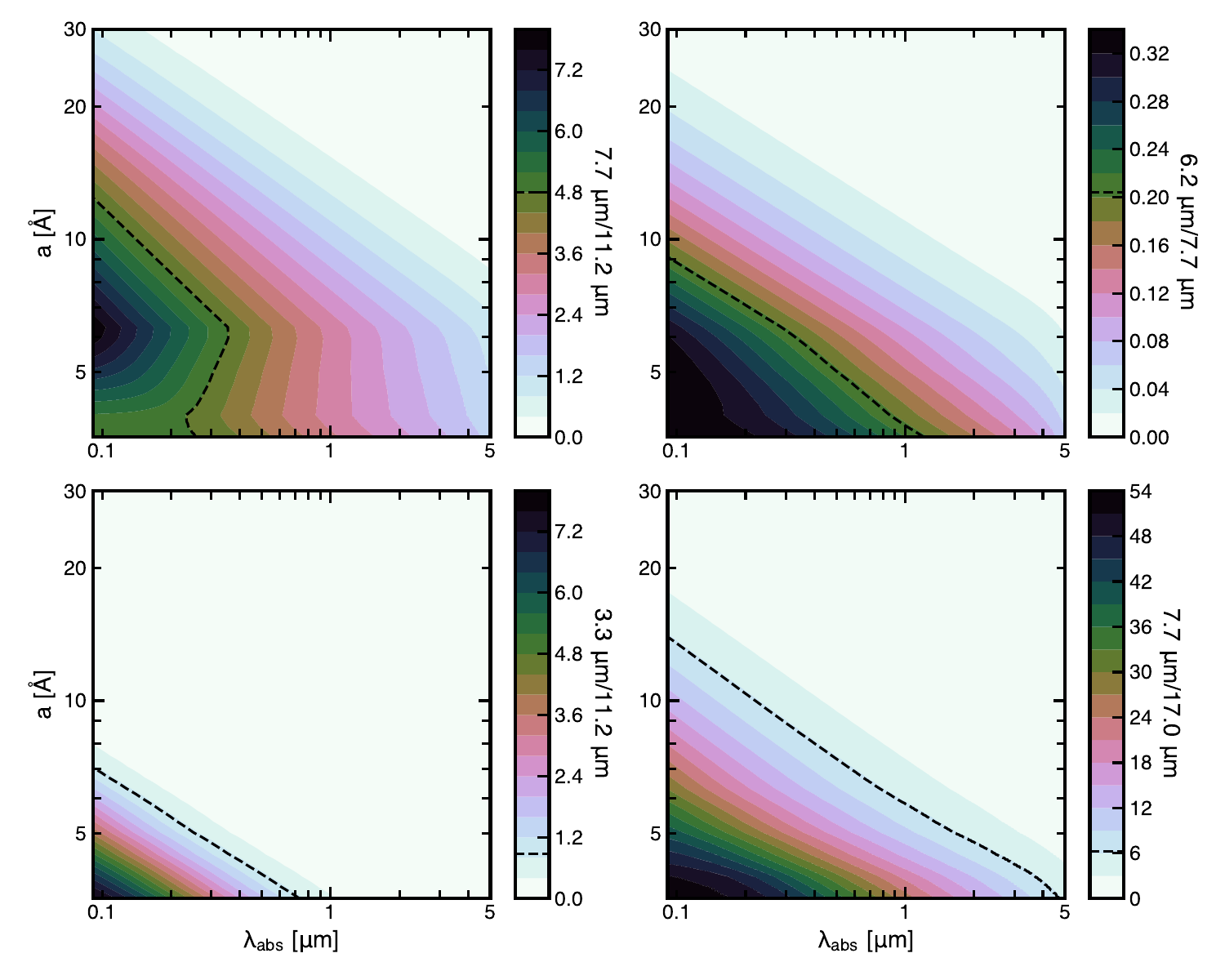}
\caption{Band ratios for the ionization-weighted basis spectra assuming the ``standard'' ionization function $f_{\rm ion}^{\rm st}\left(a\right)$ of \citet{Draine2021}, i.e., $\tilde{p}_{\lambda}=f_{\rm ion}^{\rm st}\left(a\right)\,\tilde{p}_{\lambda}^+ + (1-f_{\rm ion}^{\rm st}\left(a\right))\,\tilde{p}_{\lambda}^0$. Dashed lines highlight values for the \citet{Draine2021} models with the standard size distribution and ionization.}\label{fig:band_ratios}
\end{centering}
\end{figure*}

\begin{figure*}
\begin{centering}
\includegraphics[width=0.92\textwidth]{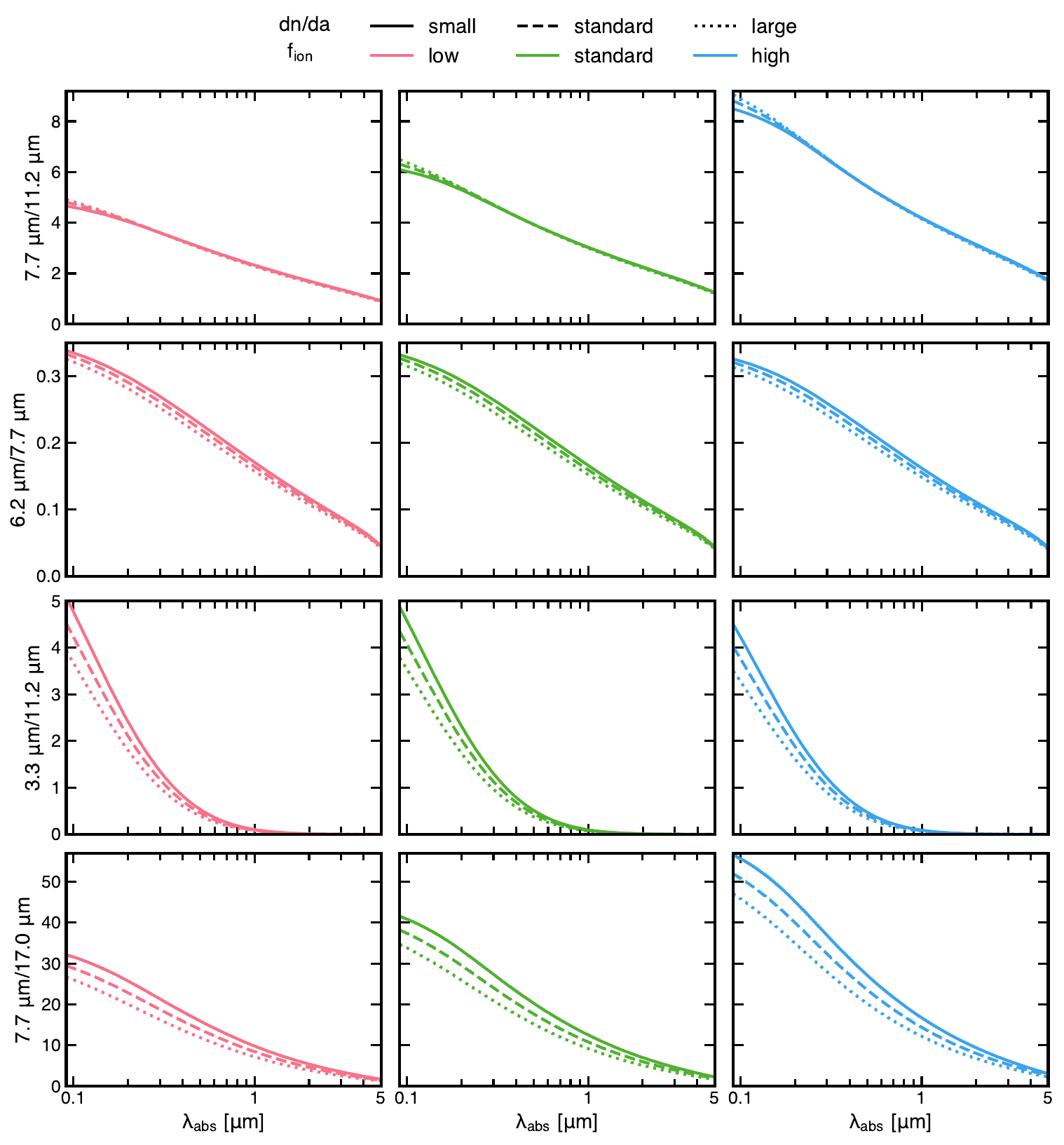}
\caption{Size- and ionization-integrated band ratios as a function of illuminating photon wavelength. In each column, we show results for the low (left), standard (middle), and high (right) ionization functions from \citet{Draine2021}. Each panel shows ratios for the small, standard, and large \citet{Draine2021} size distributions.}\label{fig:size_integrated_ratios}
\end{centering}
\end{figure*}

The relative strengths of emission features in observed PAH spectra are commonly used to characterize the underlying PAH properties, such as their size distribution and ionization function. In addition to size and ionization, the spectrum of the underlying radiation field also influences the relative strengths of these features. It is therefore challenging to untangle the simultaneous influences on band ratios to measure the physical properties of PAHs without a tool that enables fine control on radiation field properties. The single-photon spectra, $\tilde{p}_{\lambda_{\rm em}}(\lambda_{\rm abs})$, provide a way to characterize the intrinsic strengths of PAH emission features as a function of the illuminating photon energy, making it possible to disentangle radiation field effects on band ratios. In this Section, we use our model to characterize the intrinsic PAH band ratios as a function of grain size, ionization, and photon energy. 

In Figure~\ref{fig:band_ratios}, we present the ${7.7~{\rm \mu m}}/{11.2~{\rm \mu m}}$, ${6.2~{\rm \mu m}}/{7.7~{\rm \mu m}}$, ${3.3~{\rm \mu m}}/{11.3~{\rm \mu m}}$, and ${7.7~{\rm \mu m}}/{17.0~{\rm \mu m}}$ band ratios of our $\tilde{p}_{\lambda_{\rm em}}(\lambda_{\rm abs})$ basis spectra for neutral and ionized PAHs (weighted by the standard ionization function) as a function of $a$ and $\lambda_{\rm abs}$. We follow the ``clip'' method described in \citet{Draine2021} to compute the strengths of the 3.3, 6.2, 7.7, 11.2, and 17.0$~{\rm \mu m}$ emission features. This figure highlights an $a-\lambda_{\rm abs}$ degeneracy in each of the ratios, illustrating that the strength of the degeneracy varies between ratios. In particular, the ${7.7~{\rm \mu m}}/{11.2~{\rm \mu m}}$, ${3.3~{\rm \mu m}}/{11.2~{\rm \mu m}}$, and ${7.7~{\rm \mu m}}/{17.0~{\rm \mu m}}$ ratios all span a considerable range ($\gtrsim8$) across the $\lambda_{\rm abs}-a$ plane. Conversely, the ${6.2~{\rm \mu m}}/{7.7~{\rm \mu m}}$ ratio is relatively robust to changes in $a$ and $\lambda_{\rm abs}$, varying by a maximum of 0.35 throughout the parameter space. This makes the ${6.2~{\rm \mu m}}/{7.7~{\rm \mu m}}$ ratio a weak probe of the PAH size or radiation field properties, not unexpectedly since the wavelengths are so close. Although the ${3.3~{\rm \mu m}}/{11.2~{\rm \mu m}}$ ratio spans a wide range, the nonzero values of this ratio are confined to a relatively limited range of the $\lambda_{\rm abs}-a$ plane (i.e., $a\lesssim8~\angstrom$ and $\lambda_{\rm abs}\lesssim1~{\rm \mu m}$), rapidly going to zero as absorbed photon energy decreases or grain size increases. As such, the ${3.3~{\rm \mu m}}/{11.2~{\rm \mu m}}$ ratio is a useful tracer of small grains and hard radiation fields.

Although there is a degeneracy between $a$ and $\lambda_{\rm abs}$ for each of these ratios, the trend is monotonic (i.e., the ratios consistently increase with decreasing $a$ and/or $\lambda_{\rm abs}$) for the ${6.2~{\rm \mu m}}/{7.7~{\rm \mu m}}$, ${3.3~{\rm \mu m}}/{11.2~{\rm \mu m}}$, and ${7.7~{\rm \mu m}}/{17.0~{\rm \mu m}}$ ratios. However, this is not true for the ${7.7~{\rm \mu m}}/{11.7~{\rm \mu m}}$ ratio---instead it exhibits a peak at $a\sim6.5~{\angstrom}$ irrespective of absorbed photon energy.

In Figure~\ref{fig:size_integrated_ratios}, we show these same band ratios, collapsed over the $a$-axis by integrating over the various size distributions and ionization functions from \citet{Draine2021}. For band ratios that are sensitive to the size distribution (i.e., ${6.2~{\rm \mu m}}/{7.7~{\rm \mu m}}$, ${3.3~{\rm \mu m}}/{11.2~{\rm \mu m}}$, and ${7.7~{\rm \mu m}}/{17.0~{\rm \mu m}}$), there is a $\lambda_{\rm abs}$-dependency in the band ratio spread between size distributions. Additionally, since the ${3.3~{\rm \mu m}}/{11.2~{\rm \mu m}}$ ratio falls to zero beyond $\sim1~{\rm \mu m}$, it can serve as a proxy for radiation field hardness.

The trends we find for strengths of ${7.7~{\rm \mu m}}/{11.2~{\rm \mu m}}$ and ${3.3~{\rm \mu m}}/{11.2~{\rm \mu m}}$ ratios as a function of $a$ and $\lambda_{\rm abs}$ are consistent with those presented in Figure~8 of \citet{Rigopoulou:2021}. However, the non-monotonicity of the ${7.7~{\rm \mu m}}/{11.2~{\rm \mu m}}$ band strength as a function of $a$ is not captured in their work since they focus on only two grain sizes.

\section{Discussion} \label{sec:discussion}

The SPA for PAH heating and emission presents a new way to quickly generate PAH spectra for a wide range of radiation environments. As shown in Section~\ref{sec:band_ratios}, the illuminating photon energy can be a dominant driver of PAH emission feature strength. Therefore, it is crucial to consider the differences in the radiation environment when modeling PAH spectra. Our model opens up a pathway for constraining the radiation field through, e.g., Markov Chain Monte Carlo-style fitting routines, since it is relatively quick to scale the $\tilde{p}_{\lambda_{\rm em}}(\lambda_{\rm abs})$ to an input radiation field given a set of pre-computed $\tilde{p}_{\lambda_{\rm em}}(\lambda_{\rm abs})$. This framework also enables the exploration of a wider range of PAH models. Since the approach outlined in Section~\ref{sec:single-photon-approx} is independent of the choice of PAH model, it can be used with arbitrary absorption cross-section models, energy models, etc. (though the initial step of computing $\tilde{p}_{\rm \lambda_{em}}(\lambda_{\rm abs})$ does take considerable time).

With the flexibility to model PAH emission in diverse radiation environments, we can better constrain PAH properties in regions where the radiation field is expected to change significantly. A prime example is galactic outflows, an environment in which considerable variation in the radiation field is likely, but $U$ remains low enough that the SPA holds (see, e.g., \citealt{Leroy2015}). This model can be used in conjunction with existing maps of PAH emission in outflows (e.g., \citealt{Chastenet2024, Bolatto2024, Veilleux2025, Sutter2025, Lopez2025}) to untangle radiation field effects and the underlying PAH evolution. 

This framework also provides the ability to generate PAH spectral energy distributions (SEDs) from simulated data. Since our method is computationally efficient, it is suitable for use in radiative transfer modeling. This method would be straightforward to incorporate into, e.g., the \texttt{POWDERDAY} radiative transfer code \citep{Narayanan2021}, which has an explicit model for PAH evolution and heating \citep{Narayanan2023}, to generate exact PAH SEDs on the fly. Our framework may also inform the interpretation of galaxy-scale observations of PAH emission, especially where it is desirable to simultaneously constrain properties of the radiation environment, e.g., as a connection to star formation history. Observationally-derived maps of $U$ in a large sample of nearby galaxies demonstrate that average $U$ values are typically below $\sim$100 \citep{Aniano:2020, Chastenet2025}, the regime in which the SPA is a good approximation. In addition to resolved observations of PAH emission, the SPA framework could be applied to integrated observations, e.g., of the 3.3\,$\mu$m feature from SPHEREx from $0 < z < 0.5$ \citep{Cheng:2025, Zhang:2025} or PAH features across the high-redshift Universe with a mission like PRIMA \citep{Cheng:2025, Glenn:2025, Yoon:2025}.

In comparing the SPA models with those of \citet{Draine2021}, significant disagreement is evident at the locations of the narrow emission features, given the large spikes in the residuals shown in Figure~\ref{fig:residuals}. These disagreements are due to interpolation errors in the cross-section calculations of \citet{Draine2021} that cause the emission feature profiles to disagree slightly in shape and strength with the analytic $C_{\rm abs}$ profiles. This problem is most severe for the $3.3~{\rm \mu m}$ feature. Since our calculations use the analytic form of $C_{\rm abs}$, no interpolation is required. Thus, if more precise knowledge of the emission feature profiles is needed, our high-resolution calculations of the \citet{Draine2021} models may be more suitable.

The primary limitation of this model is when the SPA breaks down due to multi-photon effects, e.g., in cases where the radiation field is intense or there is a high abundance of larger PAHs. For our calculations using the standard size distribution and ionization function, we find that multi-photon effects introduce $>10\%$ effects on the spectrum in the $3-10~{\rm \mu m}$ range for $U\gtrsim100$. Since small grains remain in the single-photon limit for high radiation field intensities (as shown in Figure~\ref{fig:single_grain_spectrum}), spectra for PAH populations skewed toward smaller sizes may be valid in more intense radiation fields.

\section{Software and Data} \label{sec:code}

With the release of this paper, we make available software and data that can be used to generate SPA emission spectra \citep{RichieCode2026, RichieData2025}.\footnote{The code, instructions for its use, and documented examples are available in our online code repository, \href{https://github.com/helenarichie/pah_spec}{https://github.com/helenarichie/pah\_spec}.}\textsuperscript{,}\footnote{The data are available at \href{https://doi.org/10.7910/DVN/LUUXEJ}{https://doi.org/10.7910/DVN/LUUXEJ.}}
The code contains two main components: routines for calculating the basis spectra (Equation~\ref{eq:basisvector}) and routines for scaling the basis spectra to an input radiation field (Equation~\ref{eq:scaled_basis_vector}). 
The former enables users to generate their own basis spectra using different PAH physics models.
The code also provides functions for computing PAH energies and absorption cross-sections according to the models described in Section~\ref{sec:pah_model}.
Documented examples for generating basis and emission spectra can be found in our code repository. 
We implement the \citet{Draine2021} size distributions and ionization functions and $U=1$ mMMP radiation field as default inputs for computing the spectrum.

Because the basis spectra are time-consuming to compute, we have published a set of pre-computed $\tilde{p}_{\lambda_{\rm em}}(\lambda_{\rm abs})$  \citep{RichieData2025}, which can be quickly scaled to an arbitrary radiation field and used to create spectra for arbitrary grain size distributions and ionization functions. 
These $\tilde{p}_{\lambda_{\rm em}}(\lambda_{\rm abs})$ were computed using the PAH model described in Section~\ref{sec:pah_model}.
The basis spectra are computed for neutral and ionized PAHs ranging in size from $a=3.5$--$100~\angstrom$. 
Each grain size/ionization contains a set of $\tilde{p}_{\lambda_{\rm em}}(\lambda_{\rm abs})$ defined for each $\lambda_{\rm abs}$ in the range $0.0912-10.1~{\rm \mu m}$ with $\Delta=0.01$ ($R=100$).
They are defined over a range of $0.1\leq\lambda_{\rm em}/{\rm \mu m}\leq10^4$. 
We employ a spectral resolution of $R=2700$ in the $\lambda_{\rm em}=1$--$20~{\rm \mu m}$ range, and $R=100$ for all other wavelengths. 

\section{Conclusions} \label{sec:conclusions}

The principal conclusions of this work are as follows:

\begin{enumerate}
    \item We present a new method for generating PAH emission spectra in arbitrary radiation environments. Leveraging the single photon approximation (SPA) for PAH heating and emission, this model treats PAH spectra as a linear combination of basis spectra corresponding to individual photon absorptions at varying photon energies. 
    \item With the SPA, we reproduce spectra computed accounting for multi-photon effects \citep{Draine2021} to within $\simeq10\%$ in the $3$--$20~{\rm \mu m}$ wavelength range for radiation field strengths $U \lesssim 100$. 
    \item We find that the accuracy of the SPA is not sensitive to the radiation field spectrum over a wide range of model spectra. However, the SPA breaks down for large grains and intense radiation fields.
    \item The single photon spectra provided by this framework elucidate the degeneracies between the illuminating radiation energy and PAH properties such as size distribution and ionization (see Figures~\ref{fig:size_integrated_ratios} and \ref{fig:band_ratios}). We find the 3.3\,$\mu$m feature strength to be particularly sensitive to the hardness of the radiation field.
\end{enumerate}

We anticipate the framework developed here to be particularly useful for fitting JWST PAH emission spectra with parametric models that include variations in the spectrum of the illuminating radiation. Likewise, this formalism enables forward modeling of emission spectra given an input radiation field, e.g., from simulations. We have made all software and data products publicly available.

\begin{acknowledgments}
We thank Bruce Draine for access to and assistance with the software used in computing the \citet{Draine2021} spectra, as well as for helpful conversations. We thank Dalya Baron, Grant Donnelly, Desika Narayanan, and Karin Sandstrom for valuable discussions.

This work originated from a project of the Summer Program in Astrophysics 2025 held at the University of Virginia, and funded by the Center for Global Inquiry and Innovation, the National Science Foundation (Grant 2452494), the National Radio Astronomy Observatory (NRAO), the Kavli Foundation and the Heising-Simons Foundation. Part of this work was carried out at the Jet Propulsion Laboratory, California Institute of Technology, under a contract with the National Aeronautics and Space Administration (80NM0018D0004).
\end{acknowledgments}

\software{\texttt{pah\_spec} \citep{RichieCode2026}, Astropy \citep{2013A&A...558A..33A, 2018AJ....156..123A, 2022ApJ...935..167A}, SciPy \citep{2020SciPy-NMeth}, pandas \citep{reback2020pandas}, NumPy \citep{harris2020array}, Matplotlib \citep{Hunter:2007}}

\bibliography{sample701}{}

@article{Glenn:2025,
author = {Jason Glenn and Margaret Meixner and Charles  M. Bradford and Klaus Pontoppidan and Alexandra Pope and Tiffany Kataria and Jennifer Rocca and Elizabeth Luthman and Lee Armus and Jochem Baselmans and Cara Battersby and Alberto Bolatto and Denis Burgarella and Weibo Chen and Laure Ciesla and Peter Day and Anna Di Giorgio and Michael DiPirro and Charles Darren Dowell and Pierre Echternach and Thomas Essinger-Hileman and Marc Foote and Carlotta Gruppioni and Brandon Hensley and Thomas Henning and Willem Jellema and Matthew Johnson and Alan Kogut and Oliver Krause and James McGuire and Elisabeth Mills and Arielle Moullet and Michael Rodgers and Marc Sauvage and John  D. Smith and Rachel Somerville and Johannes Staguhn and Thomas Stevenson and Carole Tucker and Stephen Unwin and John Ziemer and Matthew Cannella and Richard Dissly},
title = {{PRIMA mission concept}},
volume = {11},
journal = {Journal of Astronomical Telescopes, Instruments, and Systems},
number = {3},
publisher = {SPIE},
pages = {031628},
year = {2025},
doi = {10.1117/1.JATIS.11.3.031628},
URL = {https://doi.org/10.1117/1.JATIS.11.3.031628}
}

@ARTICLE{Yoon:2025,
       author = {{Yoon}, Ilsang and {Hensley}, Brandon and {Lai}, Thomas S. -Y. and {Shivaei}, Irene and {Garc{\'\i}a-Bernete}, Ismael and {Donnelly}, Grant P. and {Pope}, Alexandra and {Smith}, John-David T. and {Torrey}, Paul},
        title = "{Polycyclic aromatic hydrocarbons in the high-redshift universe: prospect of the PRIMA FIRESS low-resolution spectroscopy}",
      journal = {Journal of Astronomical Telescopes, Instruments, and Systems},
         year = 2025,
        month = jul,
       volume = {11},
          eid = {031634},
        pages = {031634},
          doi = {10.1117/1.JATIS.11.3.031634},
archivePrefix = {arXiv},
       eprint = {2509.02470},
 primaryClass = {astro-ph.GA},
       adsurl = {https://ui.adsabs.harvard.edu/abs/2025JATIS..11c1634Y}
}

@ARTICLE{Zhang:2025,
       author = {{Zhang}, Edward and {Faisst}, Andreas L. and {Crill}, Brendan P. and {Inami}, Hanae and {Lai}, Thomas and {Ohyama}, Youichi and {Pyo}, Jeonghyun and {Akeson}, Rachel and {Ashby}, Matthew L.~N. and {Bock}, James J. and {Cheng}, Yun-Ting and {Chiang}, Yi-Kuan and {Cooray}, Asantha and {Dor{\'e}}, Olivier and {Feder}, Richard M. and {Kim}, Yongjung and {Lee}, Bomee and {Masters}, Daniel and {Melnick}, Gary and {Paladini}, Roberta and {Werner}, Michael W.},
        title = "{The Potential of the SPHEREx Mission for Characterizing Polycyclic Aromatic Hydrocarbon 3.3 {\ensuremath{\mu}}m Emission in Nearby Galaxies}",
      journal = {\apj},
         year = 2025,
        month = oct,
       volume = {992},
       number = {1},
          eid = {3},
        pages = {3},
          doi = {10.3847/1538-4357/adfd5a},
archivePrefix = {arXiv},
       eprint = {2503.21876},
 primaryClass = {astro-ph.GA},
       adsurl = {https://ui.adsabs.harvard.edu/abs/2025ApJ...992....3Z}
}

@ARTICLE{Cheng:2025,
       author = {{Cheng}, Yun-Ting and {Hensley}, Brandon S. and {Lai}, Thomas S. -Y.},
        title = "{Feature Intensity Mapping: Polycyclic Aromatic Hydrocarbon Emission from All Galaxies Across Cosmic Time}",
      journal = {arXiv e-prints},
         year = 2025,
        month = jun,
          eid = {arXiv:2506.13863},
        pages = {arXiv:2506.13863},
          doi = {10.48550/arXiv.2506.13863},
archivePrefix = {arXiv},
       eprint = {2506.13863},
 primaryClass = {astro-ph.CO},
       adsurl = {https://ui.adsabs.harvard.edu/abs/2025arXiv250613863C}
}

@ARTICLE{Draine:2016,
       author = {{Draine}, B.~T.},
        title = "{Graphite Revisited}",
      journal = {\apj},
         year = 2016,
        month = nov,
       volume = {831},
       number = {1},
          eid = {109},
        pages = {109},
          doi = {10.3847/0004-637X/831/1/109},
archivePrefix = {arXiv},
       eprint = {1608.02975},
 primaryClass = {astro-ph.GA},
       adsurl = {https://ui.adsabs.harvard.edu/abs/2016ApJ...831..109D}
}

@ARTICLE{Leger:1989,
       author = {{Leger}, A. and {D'Hendecourt}, L. and {Defourneau}, D.},
        title = "{Physics of IR emission by interstellar PAH molecules}",
      journal = {\aap},
         year = 1989,
        month = jun,
       volume = {216},
       number = {1-2},
        pages = {148-164},
       adsurl = {https://ui.adsabs.harvard.edu/abs/1989A&A...216..148L}
}

@ARTICLE{Whitcomb:2025,
       author = {{Whitcomb}, Cory M. and {Smith}, J. -D.~T. and {Tarantino}, Elizabeth and {Sandstrom}, Karin and {Lai}, Thomas S. -Y. and {Armus}, Lee and {Bolatto}, Alberto and {Boyer}, Martha and {Dale}, Daniel A. and {Draine}, Bruce T. and {Hensley}, Brandon S. and {Narayanan}, Desika and {Roman-Duval}, Julia and {Skillman}, Evan D.},
        title = "{The Metallicity Dependence of PAH Emission in Galaxies II: Insights from JWST/NIRCam Imaging of the Smallest Dust Grains in M101}",
      journal = {arXiv e-prints},
         year = 2025,
        month = sep,
          eid = {arXiv:2509.18347},
        pages = {arXiv:2509.18347},
          doi = {10.48550/arXiv.2509.18347},
archivePrefix = {arXiv},
       eprint = {2509.18347},
 primaryClass = {astro-ph.GA},
       adsurl = {https://ui.adsabs.harvard.edu/abs/2025arXiv250918347W}
}

@ARTICLE{Rigopoulou:2021,
       author = {{Rigopoulou}, D. and {Barale}, M. and {Clary}, D.~C. and {Shan}, X. and {Alonso-Herrero}, A. and {Garc{\'\i}a-Bernete}, I. and {Hunt}, L. and {Kerkeni}, B. and {Pereira-Santaella}, M. and {Roche}, P.~F.},
        title = "{The properties of polycyclic aromatic hydrocarbons in galaxies: constraints on PAH sizes, charge and radiation fields}",
      journal = {\mnras},
         year = 2021,
        month = jul,
       volume = {504},
       number = {4},
        pages = {5287-5300},
          doi = {10.1093/mnras/stab959},
archivePrefix = {arXiv},
       eprint = {2011.10114},
 primaryClass = {astro-ph.GA},
       adsurl = {https://ui.adsabs.harvard.edu/abs/2021MNRAS.504.5287R}
}

@ARTICLE{Desert:1986,
       author = {{Desert}, F.~X. and {Boulanger}, F. and {Shore}, S.~N.},
        title = "{Grain temperature fluctuations - A key to infrared spectra}",
      journal = {\aap},
         year = 1986,
        month = may,
       volume = {160},
       number = {2},
        pages = {295-300},
       adsurl = {https://ui.adsabs.harvard.edu/abs/1986A&A...160..295D}
}

@ARTICLE{Draine:1985,
       author = {{Draine}, B.~T. and {Anderson}, N.},
        title = "{Temperature fluctuations and infrared emission from interstellar grains.}",
      journal = {\apj},
         year = 1985,
        month = may,
       volume = {292},
        pages = {494-499},
          doi = {10.1086/163181},
       adsurl = {https://ui.adsabs.harvard.edu/abs/1985ApJ...292..494D}
}

@ARTICLE{Guhathakurta:1989,
       author = {{Guhathakurta}, P. and {Draine}, B.~T.},
        title = "{Temperature Fluctuations in the Interstellar Grains. I. Computational Method and Sublimation of Small Grains}",
      journal = {\apj},
         year = 1989,
        month = oct,
       volume = {345},
        pages = {230},
          doi = {10.1086/167899},
       adsurl = {https://ui.adsabs.harvard.edu/abs/1989ApJ...345..230G}
}

@ARTICLE{Lai:2020,
       author = {{Lai}, Thomas S. -Y. and {Smith}, J.~D.~T. and {Baba}, Shunsuke and {Spoon}, Henrik W.~W. and {Imanishi}, Masatoshi},
        title = "{All the PAHs: An AKARI-Spitzer Cross-archival Spectroscopic Survey of Aromatic Emission in Galaxies}",
      journal = {\apj},
         year = 2020,
        month = dec,
       volume = {905},
       number = {1},
          eid = {55},
        pages = {55},
          doi = {10.3847/1538-4357/abc002},
archivePrefix = {arXiv},
       eprint = {2010.05034},
 primaryClass = {astro-ph.GA},
       adsurl = {https://ui.adsabs.harvard.edu/abs/2020ApJ...905...55L}
}

@ARTICLE{Shipley:2016,
       author = {{Shipley}, Heath V. and {Papovich}, Casey and {Rieke}, George H. and {Brown}, Michael J.~I. and {Moustakas}, John},
        title = "{A New Star Formation Rate Calibration from Polycyclic Aromatic Hydrocarbon Emission Features and Application to High-redshift Galaxies}",
      journal = {\apj},
         year = 2016,
        month = feb,
       volume = {818},
       number = {1},
          eid = {60},
        pages = {60},
          doi = {10.3847/0004-637X/818/1/60},
archivePrefix = {arXiv},
       eprint = {1601.01698},
 primaryClass = {astro-ph.GA},
       adsurl = {https://ui.adsabs.harvard.edu/abs/2016ApJ...818...60S}
}

@ARTICLE{McKinney:2025,
       author = {{McKinney}, Jed and {Eleazer}, Miriam and {Pope}, Alexandra and {Sajina}, Anna and {Alberts}, Stacey and {Stone}, Meredith and {Sajkov}, Leonid and {Vanicek}, Virginia and {Kirkpatrick}, Allison and {Lai}, Thomas and {Casey}, Caitlin M. and {Armus}, Lee and {Diaz-Santos}, Tanio and {Korkus}, Andrew and {Cooper}, Olivia and {House}, Lindsay R. and {Akins}, Hollis and {Lambrides}, Erini and {Long}, Arianna and {Yan}, Lin},
        title = "{A JWST MIRI LRS Survey of 37 Massive Star-Forming Galaxies and AGN at Cosmic Noon - Overview and First Results}",
      journal = {arXiv e-prints},
         year = 2025,
        month = oct,
          eid = {arXiv:2510.07365},
        pages = {arXiv:2510.07365},
archivePrefix = {arXiv},
       eprint = {2510.07365},
 primaryClass = {astro-ph.GA},
       adsurl = {https://ui.adsabs.harvard.edu/abs/2025arXiv251007365M}
}

@ARTICLE{Aniano:2020,
       author = {{Aniano}, G. and {Draine}, B.~T. and {Hunt}, L.~K. and {Sandstrom}, K. and {Calzetti}, D. and {Kennicutt}, R.~C. and {Dale}, D.~A. and {Galametz}, M. and {Gordon}, K.~D. and {Leroy}, A.~K. and {Smith}, J. -D.~T. and {Roussel}, H. and {Sauvage}, M. and {Walter}, F. and {Armus}, L. and {Bolatto}, A.~D. and {Boquien}, M. and {Crocker}, A. and {De Looze}, I. and {Donovan Meyer}, J. and {Helou}, G. and {Hinz}, J. and {Johnson}, B.~D. and {Koda}, J. and {Miller}, A. and {Montiel}, E. and {Murphy}, E.~J. and {Rela{\~n}o}, M. and {Rix}, H. -W. and {Schinnerer}, E. and {Skibba}, R. and {Wolfire}, M.~G. and {Engelbracht}, C.~W.},
        title = "{Modeling Dust and Starlight in Galaxies Observed by Spitzer and Herschel: The KINGFISH Sample}",
      journal = {\apj},
         year = 2020,
        month = feb,
       volume = {889},
       number = {2},
          eid = {150},
        pages = {150},
          doi = {10.3847/1538-4357/ab5fdb},
archivePrefix = {arXiv},
       eprint = {1912.04914},
 primaryClass = {astro-ph.GA},
       adsurl = {https://ui.adsabs.harvard.edu/abs/2020ApJ...889..150A}
}

@ARTICLE{Whitcomb:2024,
       author = {{Whitcomb}, Cory M. and {Smith}, J. -D.~T. and {Sandstrom}, Karin and {Starkey}, Carl A. and {Donnelly}, Grant P. and {Draine}, Bruce T. and {Skillman}, Evan D. and {Dale}, Daniel A. and {Armus}, Lee and {Hensley}, Brandon S. and {Lai}, Thomas S. -Y. and {Kennicutt}, Robert C.},
        title = "{The Metallicity Dependence of PAH Emission in Galaxies. I. Insights from Deep Radial Spitzer Spectroscopy}",
      journal = {\apj},
         year = 2024,
        month = oct,
       volume = {974},
       number = {1},
          eid = {20},
        pages = {20},
          doi = {10.3847/1538-4357/ad66c8},
archivePrefix = {arXiv},
       eprint = {2405.09685},
 primaryClass = {astro-ph.GA},
       adsurl = {https://ui.adsabs.harvard.edu/abs/2024ApJ...974...20W}
}

@ARTICLE{Maragkoudakis:2020,
       author = {{Maragkoudakis}, A. and {Peeters}, E. and {Ricca}, A.},
        title = "{Probing the size and charge of polycyclic aromatic hydrocarbons}",
      journal = {\mnras},
         year = 2020,
        month = may,
       volume = {494},
       number = {1},
        pages = {642-664},
          doi = {10.1093/mnras/staa681},
archivePrefix = {arXiv},
       eprint = {2003.02823},
 primaryClass = {astro-ph.GA},
       adsurl = {https://ui.adsabs.harvard.edu/abs/2020MNRAS.494..642M}
}

@ARTICLE{2022ApJ...935..167A,
       author = {{Astropy Collaboration} and {Price-Whelan}, Adrian M. and {Lim}, Pey Lian and {Earl}, Nicholas and {Starkman}, Nathaniel and {Bradley}, Larry and {Shupe}, David L. and {Patil}, Aarya A. and {Corrales}, Lia and {Brasseur}, C.~E. and {N{\"o}the}, Maximilian and {Donath}, Axel and {Tollerud}, Erik and {Morris}, Brett M. and {Ginsburg}, Adam and {Vaher}, Eero and {Weaver}, Benjamin A. and {Tocknell}, James and {Jamieson}, William and {van Kerkwijk}, Marten H. and {Robitaille}, Thomas P. and {Merry}, Bruce and {Bachetti}, Matteo and {G{\"u}nther}, H. Moritz and {Aldcroft}, Thomas L. and {Alvarado-Montes}, Jaime A. and {Archibald}, Anne M. and {B{\'o}di}, Attila and {Bapat}, Shreyas and {Barentsen}, Geert and {Baz{\'a}n}, Juanjo and {Biswas}, Manish and {Boquien}, M{\'e}d{\'e}ric and {Burke}, D.~J. and {Cara}, Daria and {Cara}, Mihai and {Conroy}, Kyle E. and {Conseil}, Simon and {Craig}, Matthew W. and {Cross}, Robert M. and {Cruz}, Kelle L. and {D'Eugenio}, Francesco and {Dencheva}, Nadia and {Devillepoix}, Hadrien A.~R. and {Dietrich}, J{\"o}rg P. and {Eigenbrot}, Arthur Davis and {Erben}, Thomas and {Ferreira}, Leonardo and {Foreman-Mackey}, Daniel and {Fox}, Ryan and {Freij}, Nabil and {Garg}, Suyog and {Geda}, Robel and {Glattly}, Lauren and {Gondhalekar}, Yash and {Gordon}, Karl D. and {Grant}, David and {Greenfield}, Perry and {Groener}, Austen M. and {Guest}, Steve and {Gurovich}, Sebastian and {Handberg}, Rasmus and {Hart}, Akeem and {Hatfield-Dodds}, Zac and {Homeier}, Derek and {Hosseinzadeh}, Griffin and {Jenness}, Tim and {Jones}, Craig K. and {Joseph}, Prajwel and {Kalmbach}, J. Bryce and {Karamehmetoglu}, Emir and {Ka{\l}uszy{\'n}ski}, Miko{\l}aj and {Kelley}, Michael S.~P. and {Kern}, Nicholas and {Kerzendorf}, Wolfgang E. and {Koch}, Eric W. and {Kulumani}, Shankar and {Lee}, Antony and {Ly}, Chun and {Ma}, Zhiyuan and {MacBride}, Conor and {Maljaars}, Jakob M. and {Muna}, Demitri and {Murphy}, N.~A. and {Norman}, Henrik and {O'Steen}, Richard and {Oman}, Kyle A. and {Pacifici}, Camilla and {Pascual}, Sergio and {Pascual-Granado}, J. and {Patil}, Rohit R. and {Perren}, Gabriel I. and {Pickering}, Timothy E. and {Rastogi}, Tanuj and {Roulston}, Benjamin R. and {Ryan}, Daniel F. and {Rykoff}, Eli S. and {Sabater}, Jose and {Sakurikar}, Parikshit and {Salgado}, Jes{\'u}s and {Sanghi}, Aniket and {Saunders}, Nicholas and {Savchenko}, Volodymyr and {Schwardt}, Ludwig and {Seifert-Eckert}, Michael and {Shih}, Albert Y. and {Jain}, Anany Shrey and {Shukla}, Gyanendra and {Sick}, Jonathan and {Simpson}, Chris and {Singanamalla}, Sudheesh and {Singer}, Leo P. and {Singhal}, Jaladh and {Sinha}, Manodeep and {Sip{\H{o}}cz}, Brigitta M. and {Spitler}, Lee R. and {Stansby}, David and {Streicher}, Ole and {{\v{S}}umak}, Jani and {Swinbank}, John D. and {Taranu}, Dan S. and {Tewary}, Nikita and {Tremblay}, Grant R. and {de Val-Borro}, Miguel and {Van Kooten}, Samuel J. and {Vasovi{\'c}}, Zlatan and {Verma}, Shresth and {de Miranda Cardoso}, Jos{\'e} Vin{\'\i}cius and {Williams}, Peter K.~G. and {Wilson}, Tom J. and {Winkel}, Benjamin and {Wood-Vasey}, W.~M. and {Xue}, Rui and {Yoachim}, Peter and {Zhang}, Chen and {Zonca}, Andrea and {Astropy Project Contributors}},
        title = "{The Astropy Project: Sustaining and Growing a Community-oriented Open-source Project and the Latest Major Release (v5.0) of the Core Package}",
      journal = {\apj},
         year = 2022,
        month = aug,
       volume = {935},
       number = {2},
          eid = {167},
        pages = {167},
          doi = {10.3847/1538-4357/ac7c74},
archivePrefix = {arXiv},
       eprint = {2206.14220},
 primaryClass = {astro-ph.IM},
       adsurl = {https://ui.adsabs.harvard.edu/abs/2022ApJ...935..167A}
}

@ARTICLE{2018AJ....156..123A,
       author = {{Astropy Collaboration} and {Price-Whelan}, A.~M. and {Sip{\H{o}}cz}, B.~M. and {G{\"u}nther}, H.~M. and {Lim}, P.~L. and {Crawford}, S.~M. and {Conseil}, S. and {Shupe}, D.~L. and {Craig}, M.~W. and {Dencheva}, N. and {Ginsburg}, A. and {VanderPlas}, J.~T. and {Bradley}, L.~D. and {P{\'e}rez-Su{\'a}rez}, D. and {de Val-Borro}, M. and {Aldcroft}, T.~L. and {Cruz}, K.~L. and {Robitaille}, T.~P. and {Tollerud}, E.~J. and {Ardelean}, C. and {Babej}, T. and {Bach}, Y.~P. and {Bachetti}, M. and {Bakanov}, A.~V. and {Bamford}, S.~P. and {Barentsen}, G. and {Barmby}, P. and {Baumbach}, A. and {Berry}, K.~L. and {Biscani}, F. and {Boquien}, M. and {Bostroem}, K.~A. and {Bouma}, L.~G. and {Brammer}, G.~B. and {Bray}, E.~M. and {Breytenbach}, H. and {Buddelmeijer}, H. and {Burke}, D.~J. and {Calderone}, G. and {Cano Rodr{\'\i}guez}, J.~L. and {Cara}, M. and {Cardoso}, J.~V.~M. and {Cheedella}, S. and {Copin}, Y. and {Corrales}, L. and {Crichton}, D. and {D'Avella}, D. and {Deil}, C. and {Depagne}, {\'E}. and {Dietrich}, J.~P. and {Donath}, A. and {Droettboom}, M. and {Earl}, N. and {Erben}, T. and {Fabbro}, S. and {Ferreira}, L.~A. and {Finethy}, T. and {Fox}, R.~T. and {Garrison}, L.~H. and {Gibbons}, S.~L.~J. and {Goldstein}, D.~A. and {Gommers}, R. and {Greco}, J.~P. and {Greenfield}, P. and {Groener}, A.~M. and {Grollier}, F. and {Hagen}, A. and {Hirst}, P. and {Homeier}, D. and {Horton}, A.~J. and {Hosseinzadeh}, G. and {Hu}, L. and {Hunkeler}, J.~S. and {Ivezi{\'c}}, {\v{Z}}. and {Jain}, A. and {Jenness}, T. and {Kanarek}, G. and {Kendrew}, S. and {Kern}, N.~S. and {Kerzendorf}, W.~E. and {Khvalko}, A. and {King}, J. and {Kirkby}, D. and {Kulkarni}, A.~M. and {Kumar}, A. and {Lee}, A. and {Lenz}, D. and {Littlefair}, S.~P. and {Ma}, Z. and {Macleod}, D.~M. and {Mastropietro}, M. and {McCully}, C. and {Montagnac}, S. and {Morris}, B.~M. and {Mueller}, M. and {Mumford}, S.~J. and {Muna}, D. and {Murphy}, N.~A. and {Nelson}, S. and {Nguyen}, G.~H. and {Ninan}, J.~P. and {N{\"o}the}, M. and {Ogaz}, S. and {Oh}, S. and {Parejko}, J.~K. and {Parley}, N. and {Pascual}, S. and {Patil}, R. and {Patil}, A.~A. and {Plunkett}, A.~L. and {Prochaska}, J.~X. and {Rastogi}, T. and {Reddy Janga}, V. and {Sabater}, J. and {Sakurikar}, P. and {Seifert}, M. and {Sherbert}, L.~E. and {Sherwood-Taylor}, H. and {Shih}, A.~Y. and {Sick}, J. and {Silbiger}, M.~T. and {Singanamalla}, S. and {Singer}, L.~P. and {Sladen}, P.~H. and {Sooley}, K.~A. and {Sornarajah}, S. and {Streicher}, O. and {Teuben}, P. and {Thomas}, S.~W. and {Tremblay}, G.~R. and {Turner}, J.~E.~H. and {Terr{\'o}n}, V. and {van Kerkwijk}, M.~H. and {de la Vega}, A. and {Watkins}, L.~L. and {Weaver}, B.~A. and {Whitmore}, J.~B. and {Woillez}, J. and {Zabalza}, V. and {Astropy Contributors}},
        title = "{The Astropy Project: Building an Open-science Project and Status of the v2.0 Core Package}",
      journal = {\aj},
         year = 2018,
        month = sep,
       volume = {156},
       number = {3},
          eid = {123},
        pages = {123},
          doi = {10.3847/1538-3881/aabc4f},
archivePrefix = {arXiv},
       eprint = {1801.02634},
 primaryClass = {astro-ph.IM},
       adsurl = {https://ui.adsabs.harvard.edu/abs/2018AJ....156..123A}
}

@ARTICLE{2013A&A...558A..33A,
       author = {{Astropy Collaboration} and {Robitaille}, Thomas P. and
         {Tollerud}, Erik J. and {Greenfield}, Perry and {Droettboom}, Michael and
         {Bray}, Erik and {Aldcroft}, Tom and {Davis}, Matt and
         {Ginsburg}, Adam and {Price-Whelan}, Adrian M. and
         {Kerzendorf}, Wolfgang E. and {Conley}, Alexander and {Crighton}, Neil and
         {Barbary}, Kyle and {Muna}, Demitri and {Ferguson}, Henry and
         {Grollier}, Fr{\'e}d{\'e}ric and {Parikh}, Madhura M. and
         {Nair}, Prasanth H. and {Unther}, Hans M. and {Deil}, Christoph and
         {Woillez}, Julien and {Conseil}, Simon and {Kramer}, Roban and
         {Turner}, James E.~H. and {Singer}, Leo and {Fox}, Ryan and
         {Weaver}, Benjamin A. and {Zabalza}, Victor and {Edwards}, Zachary I. and
         {Azalee Bostroem}, K. and {Burke}, D.~J. and {Casey}, Andrew R. and
         {Crawford}, Steven M. and {Dencheva}, Nadia and {Ely}, Justin and
         {Jenness}, Tim and {Labrie}, Kathleen and {Lim}, Pey Lian and
         {Pierfederici}, Francesco and {Pontzen}, Andrew and {Ptak}, Andy and
         {Refsdal}, Brian and {Servillat}, Mathieu and {Streicher}, Ole},
        title = "{Astropy: A community Python package for astronomy}",
      journal = {\aap},
         year = "2013",
        month = "Oct",
       volume = {558},
          eid = {A33},
        pages = {A33},
          doi = {10.1051/0004-6361/201322068},
archivePrefix = {arXiv},
       eprint = {1307.6212},
 primaryClass = {astro-ph.IM},
       adsurl = {https://ui.adsabs.harvard.edu/abs/2013A&A...558A..33A}
}

@ARTICLE{Draine2001,
       author = {{Draine}, B.~T. and {Li}, Aigen},
        title = "{Infrared Emission from Interstellar Dust. I. Stochastic Heating of Small Grains}",
      journal = {\apj},
         year = 2001,
        month = apr,
       volume = {551},
       number = {2},
        pages = {807-824},
          doi = {10.1086/320227},
archivePrefix = {arXiv},
       eprint = {astro-ph/0011318},
 primaryClass = {astro-ph},
       adsurl = {https://ui.adsabs.harvard.edu/abs/2001ApJ...551..807D}
}

@ARTICLE{Draine2007,
       author = {{Draine}, B.~T. and {Li}, Aigen},
        title = "{Infrared Emission from Interstellar Dust. IV. The Silicate-Graphite-PAH Model in the Post-Spitzer Era}",
      journal = {\apj},
         year = 2007,
        month = mar,
       volume = {657},
       number = {2},
        pages = {810-837},
          doi = {10.1086/511055},
archivePrefix = {arXiv},
       eprint = {astro-ph/0608003},
 primaryClass = {astro-ph},
       adsurl = {https://ui.adsabs.harvard.edu/abs/2007ApJ...657..810D}
}

@ARTICLE{Krumhansl1953,
       author = {{Krumhansl}, J. and {Brooks}, H.},
        title = "{The Lattice Vibration Specific Heat of Graphite}",
      journal = {\jcp},
         year = 1953,
        month = oct,
       volume = {21},
       number = {10},
        pages = {1663-1669},
          doi = {10.1063/1.1698641},
       adsurl = {https://ui.adsabs.harvard.edu/abs/1953JChPh..21.1663K}
}

@ARTICLE{Draine2021,
       author = {{Draine}, B.~T. and {Li}, Aigen and {Hensley}, Brandon S. and {Hunt}, L.~K. and {Sandstrom}, K. and {Smith}, J. -D.~T.},
        title = "{Excitation of Polycyclic Aromatic Hydrocarbon Emission: Dependence on Size Distribution, Ionization, and Starlight Spectrum and Intensity}",
      journal = {\apj},
         year = 2021,
        month = aug,
       volume = {917},
       number = {1},
          eid = {3},
        pages = {3},
          doi = {10.3847/1538-4357/abff51},
archivePrefix = {arXiv},
       eprint = {2011.07046},
 primaryClass = {astro-ph.GA},
       adsurl = {https://ui.adsabs.harvard.edu/abs/2021ApJ...917....3D}
}

@ARTICLE{Draine2025,
       author = {{Draine}, B.~T. and {Li}, Aigen and {Hensley}, Brandon S. and {Hunt}, L.~K. and {Sandstrom}, K. and {Smith}, J. -D.~T.},
        title = "{Erratum: ``Excitation of Polycyclic Aromatic Hydrocarbon Emission: Dependence on Size Distribution, Ionization, and Starlight Spectrum and Intensity'' (2021, ApJ, 917, 3)}",
      journal = {\apj},
         year = 2025,
        month = aug,
       volume = {989},
       number = {2},
          eid = {232},
        pages = {232},
          doi = {10.3847/1538-4357/adf737},
       adsurl = {https://ui.adsabs.harvard.edu/abs/2025ApJ...989..232D}
}

@ARTICLE{Mathis1983,
       author = {{Mathis}, J.~S. and {Mezger}, P.~G. and {Panagia}, N.},
        title = "{Interstellar radiation field and dust temperatures in the diffuse interstellar medium and in giant molecular clouds}",
      journal = {\aap},
         year = 1983,
        month = nov,
       volume = {128},
        pages = {212-229},
       adsurl = {https://ui.adsabs.harvard.edu/abs/1983A&A...128..212M}
}

@BOOK{Draine2011,
       author = {{Draine}, Bruce T.},
        title = "{Physics of the Interstellar and Intergalactic Medium}",
         year = 2011,
       adsurl = {https://ui.adsabs.harvard.edu/abs/2011piim.book.....D}
}

@ARTICLE{Baron2024,
       author = {{Baron}, Dalya and {Sandstrom}, Karin M. and {Rosolowsky}, Erik and {Egorov}, Oleg V. and {Klessen}, Ralf S. and {Leroy}, Adam K. and {Boquien}, M{\'e}d{\'e}ric and {Schinnerer}, Eva and {Belfiore}, Francesco and {Groves}, Brent and {Chastenet}, J{\'e}r{\'e}my and {Dale}, Daniel A. and {Blanc}, Guillermo A. and {M{\'e}ndez-Delgado}, Jos{\'e} E. and {Koch}, Eric W. and {Grasha}, Kathryn and {Chevance}, M{\'e}lanie and {Thilker}, David A. and {Colombo}, Dario and {Williams}, Thomas G. and {Pathak}, Debosmita and {Sutter}, Jessica and {Brown}, Toby and {Wu}, John F. and {Peek}, Josh E.~G. and {Emsellem}, Eric and {Larson}, Kirsten L. and {Neumann}, Justus},
        title = "{PHANGS-ML: Dissecting Multiphase Gas and Dust in Nearby Galaxies Using Machine Learning}",
      journal = {\apj},
         year = 2024,
        month = jun,
       volume = {968},
       number = {1},
          eid = {24},
        pages = {24},
          doi = {10.3847/1538-4357/ad39e5},
archivePrefix = {arXiv},
       eprint = {2402.04330},
 primaryClass = {astro-ph.GA},
       adsurl = {https://ui.adsabs.harvard.edu/abs/2024ApJ...968...24B}
}

@ARTICLE{Baron2025,
       author = {{Baron}, Dalya and {Sandstrom}, Karin M. and {Sutter}, Jessica and {Hassani}, Hamid and {Groves}, Brent and {Leroy}, Adam K. and {Schinnerer}, Eva and {Boquien}, M{\'e}d{\'e}ric and {Brazzini}, Matilde and {Chastenet}, J{\'e}r{\'e}my and {Dale}, Daniel A. and {Egorov}, Oleg V. and {Glover}, Simon C.~O. and {Klessen}, Ralf S. and {Pathak}, Debosmita and {Rosolowsky}, Erik and {Bigiel}, Frank and {Chevance}, M{\'e}lanie and {Grasha}, Kathryn and {Hughes}, Annie and {M{\'e}ndez-Delgado}, J. Eduardo and {Pety}, J{\'e}r{\^o}me and {Williams}, Thomas G. and {Hannon}, Stephen and {Sarbadhicary}, Sumit K.},
        title = "{PHANGS-ML: The Universal Relation between PAH Band and Optical Line Ratios across Nearby Star-forming Galaxies}",
      journal = {\apj},
         year = 2025,
        month = jan,
       volume = {978},
       number = {2},
          eid = {135},
        pages = {135},
          doi = {10.3847/1538-4357/ad972a},
archivePrefix = {arXiv},
       eprint = {2410.02864},
 primaryClass = {astro-ph.GA},
       adsurl = {https://ui.adsabs.harvard.edu/abs/2025ApJ...978..135B}
}

@ARTICLE{Tielens2008,
       author = {{Tielens}, A.~G.~G.~M.},
        title = "{Interstellar polycyclic aromatic hydrocarbon molecules.}",
      journal = {\araa},
         year = 2008,
        month = sep,
       volume = {46},
        pages = {289-337},
          doi = {10.1146/annurev.astro.46.060407.145211},
       adsurl = {https://ui.adsabs.harvard.edu/abs/2008ARA&A..46..289T}
}

@ARTICLE{Li2020,
       author = {{Li}, Aigen},
        title = "{Spitzer's perspective of polycyclic aromatic hydrocarbons in galaxies}",
      journal = {Nature Astronomy},
         year = 2020,
        month = mar,
       volume = {4},
        pages = {339-351},
          doi = {10.1038/s41550-020-1051-1},
archivePrefix = {arXiv},
       eprint = {2003.10489},
 primaryClass = {astro-ph.GA},
       adsurl = {https://ui.adsabs.harvard.edu/abs/2020NatAs...4..339L}
}

@ARTICLE{Leger1984,
       author = {{Leger}, A. and {Puget}, J.~L.},
        title = "{Identification of the Unidentified Infrared Emission Features of Interstellar Dust}",
      journal = {\aap},
         year = 1984,
        month = aug,
       volume = {137},
        pages = {L5-L8},
       adsurl = {https://ui.adsabs.harvard.edu/abs/1984A&A...137L...5L}
}

@ARTICLE{Allamandola1985,
       author = {{Allamandola}, L.~J. and {Tielens}, A.~G.~G.~M. and {Barker}, J.~R.},
        title = "{Polycyclic aromatic hydrocarbons and the unidentified infrared emission bands: auto exhaust along the milky way.}",
      journal = {\apjl},
         year = 1985,
        month = mar,
       volume = {290},
        pages = {L25-L28},
          doi = {10.1086/184435},
       adsurl = {https://ui.adsabs.harvard.edu/abs/1985ApJ...290L..25A}
}

@ARTICLE{Narayanan2023,
       author = {{Narayanan}, Desika and {Smith}, J. -D.~T. and {Hensley}, Brandon S. and {Li}, Qi and {Hu}, Chia-Yu and {Sandstrom}, Karin and {Torrey}, Paul and {Vogelsberger}, Mark and {Marinacci}, Federico and {Sales}, Laura V.},
        title = "{A Framework for Modeling Polycyclic Aromatic Hydrocarbon Emission in Galaxy Evolution Simulations}",
      journal = {\apj},
         year = 2023,
        month = jul,
       volume = {951},
       number = {2},
          eid = {100},
        pages = {100},
          doi = {10.3847/1538-4357/accf8d},
archivePrefix = {arXiv},
       eprint = {2301.07136},
 primaryClass = {astro-ph.GA},
       adsurl = {https://ui.adsabs.harvard.edu/abs/2023ApJ...951..100N}
}

@ARTICLE{Narayanan2021,
       author = {{Narayanan}, Desika and {Turk}, Matthew J. and {Robitaille}, Thomas and {Kelly}, Ashley J. and {McClellan}, B. Connor and {Sharma}, Ray S. and {Garg}, Prerak and {Abruzzo}, Matthew and {Choi}, Ena and {Conroy}, Charlie and {Johnson}, Benjamin D. and {Kimock}, Benjamin and {Li}, Qi and {Lovell}, Christopher C. and {Lower}, Sidney and {Privon}, George C. and {Roberts}, Jonathan and {Sethuram}, Snigdaa and {Snyder}, Gregory F. and {Thompson}, Robert and {Wise}, John H.},
        title = "{POWDERDAY: Dust Radiative Transfer for Galaxy Simulations}",
      journal = {\apjs},
         year = 2021,
        month = jan,
       volume = {252},
       number = {1},
          eid = {12},
        pages = {12},
          doi = {10.3847/1538-4365/abc487},
archivePrefix = {arXiv},
       eprint = {2006.10757},
 primaryClass = {astro-ph.GA},
       adsurl = {https://ui.adsabs.harvard.edu/abs/2021ApJS..252...12N}
}

@ARTICLE{Leroy2015,
       author = {{Leroy}, Adam K. and {Walter}, Fabian and {Martini}, Paul and {Roussel}, H{\'e}l{\`e}ne and {Sandstrom}, Karin and {Ott}, J{\"u}rgen and {Weiss}, Axel and {Bolatto}, Alberto D. and {Schuster}, Karl and {Dessauges-Zavadsky}, Miroslava},
        title = "{The Multi-phase Cold Fountain in M82 Revealed by a Wide, Sensitive Map of the Molecular Interstellar Medium}",
      journal = {\apj},
         year = 2015,
        month = dec,
       volume = {814},
       number = {2},
          eid = {83},
        pages = {83},
          doi = {10.1088/0004-637X/814/2/83},
archivePrefix = {arXiv},
       eprint = {1509.02932},
 primaryClass = {astro-ph.GA},
       adsurl = {https://ui.adsabs.harvard.edu/abs/2015ApJ...814...83L}
}

@ARTICLE{Bolatto2024,
       author = {{Bolatto}, Alberto D. and {Levy}, Rebecca C. and {Tarantino}, Elizabeth and {Boyer}, Martha L. and {Fisher}, Deanne B. and {Cronin}, Serena A. and {Leroy}, Adam K. and {Klessen}, Ralf S. and {Smith}, J.~D. and {Berg}, Danielle A. and {B{\"o}ker}, Torsten and {Boogaard}, Leindert A. and {Ostriker}, Eve C. and {Thompson}, Todd A. and {Ott}, Juergen and {Lenki{\'c}}, Laura and {Lopez}, Laura A. and {Dale}, Daniel A. and {Veilleux}, Sylvain and {van der Werf}, Paul P. and {Glover}, Simon C.~O. and {Sandstrom}, Karin M. and {Skillman}, Evan D. and {Chisholm}, John and {Villanueva}, Vicente and {Lai}, Thomas S. -Y. and {Lopez}, Sebastian and {Mills}, Elisabeth A.~C. and {Emig}, Kimberly L. and {Armus}, Lee and {Mayya}, Divakara and {Meier}, David S. and {De Looze}, Ilse and {Herrera-Camus}, Rodrigo and {Walter}, Fabian and {Rela{\~n}o}, M{\'o}nica and {Koziol}, Hannah B. and {Marvil}, Joshua and {Jim{\'e}nez-Donaire}, Mar{\'\i}a J. and {Martini}, Paul},
        title = "{JWST Observations of Starbursts: Polycyclic Aromatic Hydrocarbon Emission at the Base of the M82 Galactic Wind}",
      journal = {\apj},
         year = 2024,
        month = may,
       volume = {967},
       number = {1},
          eid = {63},
        pages = {63},
          doi = {10.3847/1538-4357/ad33c8},
archivePrefix = {arXiv},
       eprint = {2401.16648},
 primaryClass = {astro-ph.GA},
       adsurl = {https://ui.adsabs.harvard.edu/abs/2024ApJ...967...63B}
}

@ARTICLE{Veilleux2025,
       author = {{Veilleux}, Sylvain and {Shockley}, Steven D. and {Mel{\'e}ndez}, Marcio and {Rupke}, David S.~N. and {Coil}, Alison L. and {Diamond-Stanic}, Aleksandar M. and {Geach}, James E. and {Hickox}, Ryan C. and {Moustakas}, John and {Rudnick}, Gregory H. and {Sell}, Paul H. and {Tremonti}, Christy A. and {Cha}, Hojoon},
        title = "{JWST Discovery of Warm Dust in the Circumgalactic Medium of the Makani Galaxy}",
      journal = {\apj},
         year = 2025,
        month = sep,
       volume = {990},
       number = {1},
          eid = {57},
        pages = {57},
          doi = {10.3847/1538-4357/adee91},
archivePrefix = {arXiv},
       eprint = {2507.08098},
 primaryClass = {astro-ph.GA},
       adsurl = {https://ui.adsabs.harvard.edu/abs/2025ApJ...990...57V}
}

@ARTICLE{Chastenet2024,
       author = {{Chastenet}, J{\'e}r{\'e}my and {De Looze}, Ilse and {Rela{\~n}o}, Monica and {Dale}, Daniel A. and {Williams}, Thomas G. and {Bianchi}, Simone and {Xilouris}, Emmanuel M. and {Baes}, Maarten and {Bolatto}, Alberto D. and {Boyer}, Martha L. and {Casasola}, Viviana and {Clark}, Christopher J.~R. and {Fraternali}, Filippo and {Fritz}, Jacopo and {Galliano}, Fr{\'e}d{\'e}ric and {Glover}, Simon C.~O. and {Gordon}, Karl D. and {Hirashita}, Hiroyuki and {Kennicutt}, Robert and {Nagamine}, Kentaro and {Kirchschlager}, Florian and {Klessen}, Ralf S. and {Koch}, Eric W. and {Levy}, Rebecca C. and {McCallum}, Lewis and {Madden}, Suzanne C. and {McLeod}, Anna F. and {Meidt}, Sharon E. and {Mosenkov}, Aleksandr V. and {Richie}, Helena M. and {Saintonge}, Am{\'e}lie and {Sandstrom}, Karin M. and {Schneider}, Evan E. and {Sivkova}, Evgenia E. and {Smith}, J.~D.~T. and {Smith}, Matthew W.~L. and {van der Wel}, Arjen and {Walch}, Stefanie and {Walter}, Fabian and {Wood}, Kenneth},
        title = "{JWST MIRI and NIRCam observations of NGC 891 and its circumgalactic medium}",
      journal = {\aap},
         year = 2024,
        month = oct,
       volume = {690},
          eid = {A348},
        pages = {A348},
          doi = {10.1051/0004-6361/202451033},
archivePrefix = {arXiv},
       eprint = {2408.08026},
 primaryClass = {astro-ph.GA},
       adsurl = {https://ui.adsabs.harvard.edu/abs/2024A&A...690A.348C}
}

@ARTICLE{Donnelly2024,
       author = {{Donnelly}, G.~P. and {Smith}, J.~D.~T. and {Draine}, B.~T. and {Togi}, A. and {Lai}, T.~S. -Y. and {Armus}, L. and {Dale}, D.~A. and {Charmandaris}, V.},
        title = "{The Impact of an Active Galactic Nucleus on Polycyclic Aromatic Hydrocarbon Emission in Galaxies: The Case of Ring Galaxy NGC 4138}",
      journal = {\apj},
         year = 2024,
        month = apr,
       volume = {965},
       number = {1},
          eid = {75},
        pages = {75},
          doi = {10.3847/1538-4357/ad2169},
archivePrefix = {arXiv},
       eprint = {2402.08123},
 primaryClass = {astro-ph.GA},
       adsurl = {https://ui.adsabs.harvard.edu/abs/2024ApJ...965...75D}
}

@ARTICLE{Lopez2025,
       author = {{Lopez}, Sebastian and {Ring}, Colton and {Leroy}, Adam K. and {Cronin}, Serena A. and {Bolatto}, Alberto D. and {Lopez}, Laura A. and {Villanueva}, Vicente and {Fisher}, Deanne B. and {Thompson}, Todd A. and {Armus}, Lee and {Boeker}, Torsten and {Boogaard}, Leindert A. and {Boyer}, Martha L. and {Chown}, Ryan and {Dale}, Daniel A. and {Donaghue}, Keaton and {Emig}, Kimberly and {Glover}, Simon C.~O. and {Herrera-Camus}, Rodrigo and {Klessen}, Ralf S. and {Lai}, Thomas S. -Y. and {Lenkic}, Laura and {Levy}, Rebecca C. and {Meier}, David S. and {Mills}, Elisabeth and {Ott}, Juergen and {Skillman}, Evan D. and {Smith}, J.~D.~T. and {Tarantino}, Elizabeth J. and {Veilleux}, Sylvain and {Walter}, Fabian and {van der Werf}, Paul P.},
        title = "{JWST Observations of Starbursts: PAHs Closely Trace the Cool Phase of M82's Galactic Wind}",
      journal = {arXiv e-prints},
         year = 2025,
        month = oct,
          eid = {arXiv:2510.01314},
        pages = {arXiv:2510.01314},
          doi = {10.48550/arXiv.2510.01314},
archivePrefix = {arXiv},
       eprint = {2510.01314},
 primaryClass = {astro-ph.GA},
       adsurl = {https://ui.adsabs.harvard.edu/abs/2025arXiv251001314L}
}

@ARTICLE{2020SciPy-NMeth,
  author  = {Virtanen, Pauli and Gommers, Ralf and Oliphant, Travis E. and
            Haberland, Matt and Reddy, Tyler and Cournapeau, David and
            Burovski, Evgeni and Peterson, Pearu and Weckesser, Warren and
            Bright, Jonathan and {van der Walt}, St{\'e}fan J. and
            Brett, Matthew and Wilson, Joshua and Millman, K. Jarrod and
            Mayorov, Nikolay and Nelson, Andrew R. J. and Jones, Eric and
            Kern, Robert and Larson, Eric and Carey, C J and
            Polat, {\.I}lhan and Feng, Yu and Moore, Eric W. and
            {VanderPlas}, Jake and Laxalde, Denis and Perktold, Josef and
            Cimrman, Robert and Henriksen, Ian and Quintero, E. A. and
            Harris, Charles R. and Archibald, Anne M. and
            Ribeiro, Ant{\^o}nio H. and Pedregosa, Fabian and
            {van Mulbregt}, Paul and {SciPy 1.0 Contributors}},
  title   = {{{SciPy} 1.0: Fundamental Algorithms for Scientific
            Computing in Python}},
  journal = {Nature Methods},
  year    = {2020},
  volume  = {17},
  pages   = {261--272},
  adsurl  = {https://rdcu.be/b08Wh},
  doi     = {10.1038/s41592-019-0686-2},
}

@software{reback2020pandas,
    author       = {The pandas development team},
    title        = {pandas-dev/pandas: Pandas},
    month        = feb,
    year         = 2020,
    publisher    = {Zenodo},
    version      = {latest},
    doi          = {10.5281/zenodo.3509134},
    url          = {https://doi.org/10.5281/zenodo.3509134}
}

@Article{         harris2020array,
 title         = {Array programming with {NumPy}},
 author        = {Charles R. Harris and K. Jarrod Millman and St{\'{e}}fan J.
                 van der Walt and Ralf Gommers and Pauli Virtanen and David
                 Cournapeau and Eric Wieser and Julian Taylor and Sebastian
                 Berg and Nathaniel J. Smith and Robert Kern and Matti Picus
                 and Stephan Hoyer and Marten H. van Kerkwijk and Matthew
                 Brett and Allan Haldane and Jaime Fern{\'{a}}ndez del
                 R{\'{i}}o and Mark Wiebe and Pearu Peterson and Pierre
                 G{\'{e}}rard-Marchant and Kevin Sheppard and Tyler Reddy and
                 Warren Weckesser and Hameer Abbasi and Christoph Gohlke and
                 Travis E. Oliphant},
 year          = {2020},
 month         = sep,
 journal       = {Nature},
 volume        = {585},
 number        = {7825},
 pages         = {357--362},
 doi           = {10.1038/s41586-020-2649-2},
 publisher     = {Springer Science and Business Media {LLC}},
 url           = {https://doi.org/10.1038/s41586-020-2649-2}
}

@Article{Hunter:2007,
  Author    = {Hunter, J. D.},
  Title     = {Matplotlib: A 2D graphics environment},
  Journal   = {Computing in Science \& Engineering},
  Volume    = {9},
  Number    = {3},
  Pages     = {90--95},
  publisher = {IEEE COMPUTER SOC},
  doi       = {10.1109/MCSE.2007.55},
  year      = 2007
}

@ARTICLE{Chastenet2025,
       author = {{Chastenet}, J{\'e}r{\'e}my and {Sandstrom}, Karin and {Leroy}, Adam K. and {Bot}, Caroline and {Chiang}, I-Da and {Chown}, Ryan and {Gordon}, Karl D. and {Koch}, Eric W. and {Roussel}, H{\'e}l{\`e}ne and {Sutter}, Jessica and {Williams}, Thomas G.},
        title = "{The Resolved Behavior of Dust Mass, Polycyclic Aromatic Hydrocarbon Fraction, and Radiation Field in {\ensuremath{\sim}}800 Nearby Galaxies}",
      journal = {\apjs},
         year = 2025,
        month = jan,
       volume = {276},
       number = {1},
          eid = {2},
        pages = {2},
          doi = {10.3847/1538-4365/ad8a5c},
archivePrefix = {arXiv},
       eprint = {2410.03835},
 primaryClass = {astro-ph.GA},
       adsurl = {https://ui.adsabs.harvard.edu/abs/2025ApJS..276....2C}
}

@ARTICLE{Sutter2025,
       author = {{Sutter}, Jessica and {Sandstrom}, Karin and {Chown}, Ryan and {Egorov}, Oleg and {Leroy}, Adam K. and {Chastenet}, J{\'e}r{\'e}my and {Bolatto}, Alberto D. and {Williams}, Thomas G. and {Dale}, Daniel A. and {Amiri}, Amirnezam and {Boquien}, M{\'e}d{\'e}ric and {Cao}, Yixian and {Dlamini}, Simthembile and {Emsellem}, {\'E}ric and {Pan}, Hsi-An and {Pathak}, Debosmita and {Kim}, Hwihyun and {Klessen}, Ralf S. and {Koziol}, Hannah and {Rosolowsky}, Erik and {Sarbadhicary}, Sumit K. and {Schinnerer}, Eva and {Thilker}, David A. and {{\'U}beda}, Leonardo and {Weinbeck}, Tony},
        title = "{Characterization of Two Cool Galaxy Outflow Candidates Using Mid-infrared Emission from Polycyclic Aromatic Hydrocarbons}",
      journal = {\apjl},
         year = 2025,
        month = oct,
       volume = {992},
       number = {1},
          eid = {L7},
        pages = {L7},
          doi = {10.3847/2041-8213/ae08b7},
archivePrefix = {arXiv},
       eprint = {2509.12058},
 primaryClass = {astro-ph.GA},
       adsurl = {https://ui.adsabs.harvard.edu/abs/2025ApJ...992L...7S}
}

@misc{RichieData2025,
author = {Richie, Helena},
publisher = {Harvard Dataverse},
title = {{PAH Emission Spectra and Band Ratios for Arbitrary Radiation Fields with the Single Photon Approximation}},
UNF = {UNF:6:7kmKSpJpj7UlWXBO04gLWw==},
year = {2025},
version = {V2},
doi = {10.7910/DVN/LUUXEJ},
url = {https://doi.org/10.7910/DVN/LUUXEJ}
}

@software{RichieCode2026,
  author       = {Richie, Helena},
  title        = {pah\_spec},
  month        = mar,
  year         = 2026,
  publisher    = {Zenodo},
  doi          = {10.5281/zenodo.18891079},
  url          = {https://doi.org/10.5281/zenodo.18891079},
}

@misc{DraineData2021,
author = {Draine, B. T. and Li, Aigen and Hensley, Brandon S. and Hunt, L. K. and Sandstrom, K. and Smith, J.-D. T},
publisher = {Harvard Dataverse},
title = {{PAH emission: Dependence on Starlight Spectrum, Intensity, PAH Size Distribution, and PAH Ionization}},
year = {2021},
version = {V1},
doi = {10.7910/DVN/LPUHIQ},
url = {https://doi.org/10.7910/DVN/LPUHIQ}
}
\bibliographystyle{aasjournalv7}

\end{document}